\documentclass[a4paper,12pt, epsfig]{article}
\pdfoutput=1
\usepackage{epsfig}
\usepackage{epstopdf}
\usepackage{graphicx}
\usepackage{ifthen}
\usepackage{feynmp}
\usepackage{amsmath}
\usepackage[psamsfonts]{amssymb}
\usepackage{euscript}

\usepackage{latexsym}
\usepackage[arrow,matrix,curve]{xy}

\newskip\humongous \humongous=0pt plus 1000pt minus 1000pt

\newif\ifdtup

\def\,{\hspace{-.1cm}}
\def\hsp{,\hspace{.7cm}}

\def\tf {\tilde{f}}
\def\fjl {\Phi\left(\frac{n}{q}j,\frac{n}{q} l\right)}
\def\F#1#2{\Phi\left(\frac{n}{q}{#1},\frac{n}{q}{#2}\right)}
\def\fc#1#2 {\frac{n}{q}#1\frac{n}{q}#2}

\def\e#1{\langle#1\rangle_0}
\def\ee#1{\left\langle#1\right\rangle}
\def\di#1{\left(\vcenter{\xymatrix{#1}}\right)}

\def\Os{|\Omega\rangle}
\def\mo#1{\int\frac{d{#1}}{2\pi}}
\renewcommand{\theequation}{\arabic{section}.\arabic{equation}}
\renewcommand{\(}{\begin{equation}}
\renewcommand{\)}{end{equation} \vspace{-.05in}\linebreak}

\newcounter{saveeqn}
\newcounter{savealpheqn}

\newcommand{\alpheqn}{\setcounter{saveeqn}{\value{equation}}%
  \stepcounter{saveeqn}\setcounter{equation}{0}%
  \renewcommand{\theequation}{\mbox{\arabic{section}.\arabic{saveeqn}
\alph{equation}}}
  \renewcommand{\)}{\end{equation}}}
\def\part#1{\frac{\partial}{\partial{#1}}}%
\def\group#1{\refstepcounter{equation}\setcounter{saveeqn}
 {\value{equation}}%
  \label{#1}\setcounter{equation}{0}%
\renewcommand{\theequation}{\mbox{\arabic{section}.\arabic{saveeqn}
\alph{equation}}}
  \renewcommand{\)}{\end{equation}}}
\newcommand{\reseteqn}{\setcounter{equation}{\value{saveeqn}}%
  \renewcommand{\theequation}{\arabic{section}.\arabic{equation}}%
  \renewcommand{\)}{\end{equation}}}

\newcommand{\aalpheqn}{\setcounter{saveeqn}{\value{equation}}%
  \stepcounter{saveeqn}\setcounter{equation}{0}%
  \renewcommand{\theequation}{\mbox{
        \Alph{subsection}.\arabic{saveeqn}\alph{equation}}}
   \renewcommand{\)}{\end{equation}}}
\newcommand{\areseteqn}{\setcounter{equation}{\value{saveeqn}}%
  \renewcommand{\theequation}{\Alph{subsection}.\arabic{equation}}%
  \renewcommand{\)}{\end{equation}}}

\renewcommand{\thefootnote}{\alph{footnote}}
\renewcommand{\(}{\begin{equation}}
\renewcommand{\)}{\end{equation}}
\newcommand{\ba}{\begin{eqnarray}}
\newcommand{\ea}{\end{eqnarray}}

\newcommand{\bp}{\mathop{\vtop{\ialign{##\crcr
   $\hfil\displaystyle{}\hfil$\crcr\noalign{\kern-13pt\nointerlineskip}
   \BIG{(}\hskip0pt\crcr\noalign{\kern3pt}}}}}
\newcommand{\cbp}{\mathop{\vtop{\ialign{##\crcr
   $\hfil\displaystyle{}\hfil$\crcr\noalign{\kern-13pt\nointerlineskip}
   \BIG{)}\hskip0pt\crcr\noalign{\kern3pt}}}}}
\newcommand{\pa}{\mathop{\vtop{\ialign{##\crcr
    
$\hfil\displaystyle{\oplus}\hfil$\crcr\noalign{\kern+1pt\nointerlineskip 
}
   \hspace{.08in}$^{\alpha=0}$\hskip6pt\crcr\noalign{\kern3pt}}}}}
\renewcommand{\hsp}{,\hspace{.3in}}

\newcommand{\Z}{\ensuremath{\mathbb Z}}

\catcode`\@=11
\def\vereq#1#2{\lower3pt\vbox{\baselineskip1.5pt \lineskip1.5pt
\ialign{$\m@th#1\hfill##\hfil$\crcr#2\crcr\sim\crcr}}}
\catcode`\@=12

\renewcommand{\(}{\begin{equation}}
\renewcommand{\)}{\end{equation}}

\newcommand{\beas}{\begin{eqnarray*}}
\newcommand{\eeas}{\end{eqnarray*}}

\newcommand{\bquo}{\begin{quote}}
\newcommand{\enqu}{\end{quote}}

\newcommand{\C}{{\mathbb C}}
\newcommand{\cp}{{\mathrm{\mathbb CP}}}
\newcommand{\R}{{\mathbb R}}
\renewcommand{\Z}{{\mathbb Z}}

\newcommand{\beq}{\begin{equation}}
\newcommand{\eeq}{\end{equation}}
\newcommand{\bea}{\begin{eqnarray}}
\newcommand{\eea}{\end{eqnarray}}

\newskip\humongous \humongous=0pt plus 1000pt minus 1000pt

\newif\ifdtup

\catcode`\@=11

\@addtoreset{equation}{section}

\def\@normalsize{\@setsize\normalsize{15pt}\xiipt\@xiipt
\abovedisplayskip 14pt plus3pt minus3pt%
\belowdisplayskip \abovedisplayskip
\abovedisplayshortskip \z@ plus3pt%
\belowdisplayshortskip 7pt plus3.5pt minus0pt}

\def\small{\@setsize\small{13.6pt}\xipt\@xipt
\abovedisplayskip 13pt plus3pt minus3pt%
\belowdisplayskip \abovedisplayskip
\abovedisplayshortskip \z@ plus3pt%
\belowdisplayshortskip 7pt plus3.5pt minus0pt
\def\@listi{\parsep 4.5pt plus 2pt minus 1pt
      \itemsep \parsep
      \topsep 9pt plus 3pt minus 3pt}}

\relax

\catcode`@=12

\catcode`\@=11

\def\section{\@startsection{section}{1}{\z@}{3.5ex plus 1ex minus  .2ex}{2.3ex plus .2ex}{\large\bf}}

\def\thesection{\arabic{section}}
\def\thesubsection{\arabic{section}.\arabic{subsection}}

\def\appendix{\setcounter{section}{0}
 \def\thesection{Appendix \Alph{section}}
 \def\thesubsection{\Alph{section}.\arabic{subsection}}
 \def\theequation{\Alph{section}.\arabic{equation}}}
\renewcommand{\theequation}{\arabic{section}.\arabic{equation}}

\begin{document}
\def\thefootnote{\fnsymbol{footnote}}
\def\thetitle{Entangled Neutrino States in a Toy Model QFT}
\def\autone{Jarah Evslin}
\def\affa{Institute of Modern Physics, NanChangLu 509, Lanzhou 730000, China}
\def\affb{University of the Chinese Academy of Sciences, YuQuanLu 19A, Beijing 100049, China}

\begin{center}
{\large {\bf \thetitle}}

\bigskip

\bigskip

{\large \noindent  \autone{${}^{1,2}$}\footnote{jarah@impcas.ac.cn}, Hosam Mohammed{${}^{2,1}$}\footnote{hosam@impcas.ac.cn},Emilio Ciuffoli{${}^{1}$}\footnote{ciuffoli@impcas.ac.cn},Yao Zhou{${}^{2,1}$}\footnote{yaozhou@impcas.ac.cn}} 

\vskip.7cm

1) \affa\\
2) \affb\\

\end{center}

\begin{abstract}
\noindent
It has been claimed that wave packets must be covariant and also that decohered neutrino oscillations are always revived during measurement.  These conjectures are supported by general arguments which are not specific to the electroweak theory, and so if they are true for neutrinos they will also be true for simplified models.  In this paper we produce such a simplified model in which the neutrino wave function, including its entanglement with the source particle and the environment, can be calculated explicitly in quantum field theory.  It exhibits neutrino oscillation, which is reduced at late times by decoherence due to interactions of the source with the environment.  One simple lesson from this model is that only the difference between the environmental interactions before and after neutrino emission can reduce the amplitude of neutrino oscillations.   The model will be used to test the conjectures in a companion paper.

\end{abstract}

%
\setcounter{footnote}{0}
\renewcommand{\thefootnote}{\arabic{footnote}}




\section{Introduction}

Reactor neutrino experiments report lower values of $\theta_{13}$ than accelerator experiments.  It is customary to reduce this tension by assuming the normal hierarchy and a value of the CP-violating phase $\delta$ near $270^\circ$.  This increases the expected appearance signal at accelerator experiments, allowing the small $\theta_{13}$ mixing reported by reactor experiments to produce almost as many electron (anti)neutrinos as are observed at muon (anti)neutrino beams.   But there is another logically consistent possibility.  The reactor neutrinos have lower energy, and so are expected to be more prone to decoherence than accelerator neutrinos \cite{boya2011}.  Indeed no decoherence is expected in the case of accelerator neutrinos \cite{accdec}.  In this case the reactor neutrino measurement of $\theta_{13}$, based on an analysis with no decoherence, is underestimated and the evidence for the normal hierarchy and maximal CP-violation is weakened.   Furthermore, the degradation of the signal observed by JUNO would be considerable \cite{steven}.  This possibility has been rejected by the Daya Bay collaboration~\cite{dayadec}.  However their study relied upon a neutrino wave packet model.

\subsection{Wave Packet Models of Neutrinos}

The traditional view of decoherence in neutrino oscillations comes from the quantum mechanical wave packet model.  Here neutrinos are produced as a flavor eigenstate wave packet, localized in space and time.  The lighter mass eigenstate travels faster than the others and so the wave packets corresponding to different mass eigenstates spatially separate after travelling a distance called the coherence length.  This separation leads to decoherence and therefore a decrease in amplitude of neutrino oscillations.  The spatially separated mass eigenstates may nonetheless be coherently summed by the detector if the detector has a sufficiently long coherence time, leading to a restoration of neutrino oscillations \cite{revival}.  The coherence length clearly depends on the spatial size of the wave packet, which is a parameter in such models.  It has long been recognized \cite{nuss76} that this spatial size is determined by interactions of the neutrino source particles with the environment.  Usually order of magnitude arguments are used to estimate this parameter \cite{nuss76,wilczek,rich,boriserr}, and the result is substituted into the model.

In quantum mechanics, neutrino wave packets are created by hand.  In quantum field theory (QFT) they are created consistently from electroweak interactions.  Consistent QFT treatments necessarily create neutrinos entangled to their source particles, such as unstable nuclei or mesons, and also to charged leptons which are created simultaneously.  We will refer to all of the particles involved in the interaction which produced the neutrino as source particles, including the charged leptons.  Again in this case the environment plays a role.  As noted, for example, in Ref.~\cite{giunti2012} the interactions of the source particles with the environment disentangle the neutrino from the rest of the state and so allow its treatment as a wave packet.  This disentanglement is caused by environmental interactions which effectively measure the source particles \cite{zurek}.  It is customary in QFT treatments to apply this interaction by simply projecting the entangled state onto a subsector of the Hilbert space in which the source particles have some definite position or momentum wave function, as if they were actually measured.  With the positions of the source particles specified, one can determine a space time region in which the neutrino is created and so the neutrino is again in a localized, flavor eigenstate wave packet.  Now, just like the quantum mechanical case, the different mass eigenstates travel at different speeds and so separate, leading to decoherence.

Quite a different QFT treatment appeared in Ref.~\cite{mcgreevy}.  Here the different neutrino mass eigenstates were not forcibly created in the same time window.  Of course modern neutrino experiments measure neutrinos in a fixed time window, in flavor eigenstates.  Therefore the fact that lighter neutrinos travel faster and the travel distance is fixed implies that the lighter mass eigenstates are emitted after the heavy mass eigenstates.  So instead of wave packet separation, here the wave packets coelesce, and no decoherence was reported by the authors.

How could QFT produce two such phenomenologically distinct paradigms?  In the first case, environmental interactions were imposed by hand, with a simple projection.  In the second case, environmental interactions were not included at all.  

\subsection{Wave Packets from Entanglement}

It is our goal to understand when the wave packet treatment of neutrinos is and is not reliable, and to understand how to calculate the wave packet size.  We will do this via a first principles, consistent calculation in QFT.   Papers on QFT treatments of neutrinos generally calculate the S matrix for neutrino creation and detection, which is the amplitude for the creation of a given state in the asymptotic future, long after the neutrino has been absorbed.  However we are interested in the state of the neutrino itself, and so are interested in intermediate states.  Such information can not be directly obtained from the S matrix.   It is accessible in the Schrodinger picture of QFT, in which operators are time-independent and states evolve via the action of the Hamiltonian operator.  An experiment begins with a source state entangled with the environment and the Hamiltonian evolves this initial state into the future.  This evolution creates neutrinos.  

As was noted in Ref.~\cite{mcgreevy}, it is true that different neutrino mass eigenstates may be created at different times.  Indeed, evolving the state of a ${}^{235}$U nucleus for one year in the Schrodinger picture, neutrinos may be emitted at any time during the year and so the neutrino wave function extends for one light year.  It is certainly not a localized wave packet.  In the calculation of matrix elements, one must sum over each mass eigenstate and separately integrate the interaction times over the entire year.

Now the key question is, whether at a fixed time the different mass eigenstates contribute coherently to matrix elements.  If they do, one expects to observe neutrino oscillations, if they do not, these oscillations will be damped.  Measurements occur in a flavor basis and, in modern experiments, at a reasonably well-determined time.  Therefore contributions to the relevant amplitudes come from states in which the different mass eigenstates are localized in space time at detection, meaning that the lighter neutrino was emitted later, again in agreement with \cite{mcgreevy}.

However for a coherent summation of neutrino mass eigenstates it is not sufficient that they spatially overlap.  The entire final states must agree, including the source particles and the environment.  In other words, if the state is
\beq
|\psi\rangle=|E_1\rangle\otimes|\nu_1\rangle+|E_2\rangle\otimes|\nu_2\rangle
\eeq
where $|E_i\rangle$ are the environment plus source particles part of the state and $|\nu_i\rangle$ are the neutrino mass eigenstates, then the summation is fully coherent only if the $|E_i\rangle$ are equal up to a phase.  This condition is the origin of decoherence.  The fact that the lighter neutrino was emitted later means that the source particles interacted differently with the environment, for example the unstable particle had more time to interact while the product particles had less.  This necessarily implies that the environment part of the state will be different in the case of each mass eigenstate.  The bigger the difference in mass or the further the neutrino has traveled, the bigger the difference in time between the emissions of the different mass eigenstates and so the bigger the decoherence.   

The conclusion is that while Ref.~\cite{mcgreevy} is correct that the times of the emissions of the various mass eigenstates need not agree, nonetheless if the difference exceeds some threshold then coherence will be lost.  We claim that this threshold should be interpreted as the wave packet size in the wave packet model.  In this case, decoherence will correspond to the spatial separation of the wave packets.  However it is not obvious that long measurements may now restore coherence as in Ref.~\cite{revival}.

\subsection{Our Approach}

For the questions of interest, concerning neutrino oscillations, wave packets, and decoherence, the details of the electroweak interactions do not play any essential role.  Therefore, we will work in the simplest toy model which has the features of interest, a scalar field theory in 1+1 dimensions.  Here we can, in the Schrodinger picture of QFT, numerically evolve the full entangled state to any desired moment in time to understand it.   Thus our approach is similar to that of Ref.~\cite{cgl} but including environmental interactions.  To simplify the situation yet further, we will not consider measurements of the neutrinos.  Therefore our final states will be the neutrinos themselves and we will calculate transition amplitudes and transition probabilities from states with no neutrinos to states with a neutrino.  We will see that these probabilities already have a rich phenomenology of oscillations and decoherence.  Of course it means that we cannot tell whether coherence can be revived through measurement, however we feel that a robust study of coherence revival via measurement requires a characterization of the coherence before measurement, which our method provides.

We do not model interactions with the environment by projecting on to a definite state for the environment and the source particles.   Instead all particles are consistently evolved in the Schrodinger picture of QFT.  In the calculation of probabilities, the distinct environment and source final states are incoherently summed.  


The phenomenology of wave packet models includes several potentially interesting effects, such as the revival of oscillations ruined by docoherence via long measurements in Ref.~\cite{revival}.  In \cite{mcdonald} it was asserted that, presumably as a result of revival, decoherence is unobservable in neutrino oscillation experiments.  Another claim \cite{naumov1,naumov2} is that neutrino wave functions are always ``{\it{covariant wavepackets}}."  This means that they depend on the momentum only via Lorentz scalars.   The covariant wave packet hypothesis was assumed in the experimental analysis of decoherence at Daya Bay \cite{dayadec}.   We believe that our QFT approach will allow a robust test of these claims.


Our study has three advantages over most quantum field theory (QFT) approaches to neutrino oscillations and decoherence.  First, we calculate the full, entangled state consisting of the source, the neutrinos and the environment\footnote{The key role played by the entanglement of the neutrino and the source particles in a QFT treatment has been stressed in Ref.~\cite{cgl}.  In Ref.~\cite{akqft} it is claimed that the full entangled QFT treatment leads to the same amplitudes as a wave packet treatment.  However neither study included interactions of the source with the environment.}  at arbitrary times and not just the asymptotic S-matrix.  This will allow a robust test of the covariant wave packet proposal.  Second, we explicitly consider interactions between the source and the environment\footnote{Such interactions were included in Ref.~\cite{akmoss} by including a phenomenological smearing of energies.  We instead consistently treat the interactions in QFT.}.  Third, we integrate our transition probability over the possible final states of the source and the environment.  It is this integration which leads to decoherence, reducing the amplitude of neutrino oscillations in the transition probability.

Perhaps one of the most serious attempts at the determination of the wave packet size, in the case of solar neutrinos, was Ref.~\cite{nuss76}.  Unlike later estimates, it includes an estimate of the phase angle variation resulting from each interaction instead of merely assuming that an interaction automatically results in decoherence.  However, in the case of reactor neutrinos, unlike solar neutrinos, the source nuclei are large and so the Coulomb interactions in some cases are hardly affected by a beta decay.  We will see in our example that the decoherence is not determined by the total phase induced by an interaction, but rather by the difference in the phase that would be acquired before and after the beta decay.  This difference, in the case of reactor neutrinos, may be one or two orders of magnitude smaller than the total phase, and thus the wave packet size may be expected to be an order or magnitude or two larger than may be expected by simply adapting the argument of Ref.~\cite{nuss76} to the case of reactor neutrinos.  This is one immediate lesson that may be drawn from our simple model.

We begin in Sec.~\ref{classsez} with a simplified model in which the neutrinos are created from a classical source.  This model exhibits oscillations.  However the neutrinos are always off-shell and also, because the source is classical, it cannot be entangled with the environment and so there is no decoherence.  Next in Sec.~\ref{modsez} we introduce our full model.  We include both source fields and also environment states.  Our analysis of this model is presented in Sec.~\ref{ressez}.

\section{Warm Up: A Classical Source} \label{classsez}

\subsection{The Model, Fields and States}

We do not believe that spin plays a key role in a qualitative understanding of decoherence in neutrino oscillations.  Therefore our model will involve only real scalar fields.  Similarly, we will restrict our attention to one space and one time dimension.  So long as our fields are massive, this assumption leads to only a modest reduction in computational complexity.  Finally, as our most significant assumption, we will consider one-body and two-body decays instead of three-body decays.   Therefore the scalar fields which we will call ``neutrinos" will carry no conserved lepton charge.  Nonetheless we will introduce two flavors of neutrinos, so that there will be oscillations.  

The neutrinos in our model are described by the canonical real scalar fields
\beq
\psi_i(x)=\int\frac{dp}{2\pi}\frac{1}{\sqrt{2\omega_i(p)}}\left(a_{i,-p}+a^\dagger_{i,p}\right)e^{-ipx}\hsp \omega_i(p)=\sqrt{m_i^2+p^2}
\eeq
where the index $i$ labels the mass eigenstates $\psi_1$ and $\psi_2$.  The conjugate momenta are
\beq
\pi_i(x)=-i\int\frac{dp}{2\pi}\sqrt{\frac{\omega_i(p)}{2}}\left(a_{i,-p}-a^\dagger_{i,p}\right)e^{-ipx} .
\eeq
We always work in the Schrodinger picture, so all operators such as fields and their conjugate momenta are time-independent.

The Hamiltonian will be decomposed into a free and interaction term
\beq
H=H_0+H_I\hsp H_0=\int dx \mathcal{H}_0(x)\hsp H_I=\int dx \mathcal{H}_I
\eeq
where $\mathcal{H}_0$ is the free real scalar field Hamiltonian density\footnote{While the Hamiltonian $H$ could be rewritten as a free Hamiltonian via a momentum-dependent coordinate transformation, such a transformation would not be convenient for our purposes as we will consider states in the $n$-particle Fock space of $H_0$.}
\bea
\mathcal{H}_0(x)&=&\frac{1}{2}\sum_{i=1}^2:\left(\pi_i(x)^2+\left(\partial_x\psi_i(x)\right)^2+m_i^2\psi_i(x)^2\right):\nonumber\\
H_0&=&\int dx \mathcal{H}_0(x)=\sum_{i=1}^2\int \frac{dp}{2\pi}\omega_i(p)a^\dagger_{i,p}a_{i,p}.
\eea
The interaction Hamiltonian describes neutrino creation by a classical source of size $1/(2\sqrt\alpha)$
\bea
\mathcal{H}_I(x)&=&e^{-\alpha x^2}\sum_{i=1}^2 \psi_i(x)\nonumber\\
H_I&=&\int dx \mathcal{H}_I(x)=\sqrt\frac{\pi}{\alpha}\sum_{i=1}^2\int \frac{dp}{2\pi}\frac{e^{-\frac{p^2}{4\alpha}}}{\sqrt{2\omega_i(p)}}\left(a_{i,-p}+a^\dagger_{i,p}\right).
\eea
Observe that neutrinos are created not in a mass eigenstate $\psi_i$, but rather in the superposition $\psi_1+\psi_2$ which plays the role of a flavor eigenstate in our model.

Let $\Os$ and $|i,p\rangle$ be respectively the ground state and one neutrino states of the free Hamiltonian $H_0$
\beq
a_{i,p}\Os=0\hsp
|i,p\rangle=a^\dagger_{i,p}\Os.
\eeq
The states $|i,p\rangle$ provide an orthogonal basis for the 1-particle states.  

In practice one is interested in the measurement of a neutrino at a particular position $x$.  While it is straightforward to define an orthogonal position basis for the 1-particle states, this does not reflect the basis in which neutrinos are usually measured in modern experiments.  Usually one measures both a neutrino's momentum and also position.  Clearly the uncertainty principle implies that these are each measured with a finite resolution.  Let $\sigma$ be the momentum resolution of a given detector.  For simplicity, we will consider a detector which is only sensitive to neutrinos of momentum $p_0$, although this can easily be generalized to a multichannel detector.  Then the relevant basis of 1-neutrino states will be
\beq
|i,x\rangle=\mo{p} e^{-ipx}e^{-\frac{\left(p-p_0\right)^2}{2\sigma^2}}|i,p\rangle .
\eeq
Note that while these states do form a basis for the 1-neutrino sector of the Hilbert space, they are not orthogonal
\beq
\langle i,x|j,y\rangle=\sqrt{\pi}\sigma e^{-\sigma^2(x-y)^2/4}. \label{prodeq}
\eeq

\subsection{Evolution}

To calculate the evolution of this system, we will need to know how the Hamiltonian acts on the various states.  In terrestrial neutrino experiments, multineutrino processes are too suppressed to be relevant.  Thus we will be interested only in evolution involving a single power of $H_I$ and only in 0-neutrino and 1-neutrino states.  The action of the Hamiltonian on such states is easily calculated
\beq
H_0\Os=0\hsp H_I\Os=\sqrt\frac{\pi}{\alpha}\sum_{i=1}^2\int \frac{dp}{2\pi}\frac{e^{-\frac{p^2}{4\alpha}}}{\sqrt{2\omega_i(p)}}|i,p\rangle\hsp H_0|i,p\rangle=\omega_i(p)|i,p\rangle .
\eeq
$H_I|i,p\rangle$ will not arise in the calculation below at first order in $H_I$.

Projecting onto the 1-neutrino sector of the Hilbert space, we then find the evolution of the ground state to an arbitrary time $t$
\bea
|t\rangle=e^{-iHt}\Os&=&\sum_{j=0}^\infty \frac{(-iHt)^j}{j!}\Os\supset\sum_{j=1}^\infty \frac{(-it)^j}{j!}H_0^{j-1}H_I\Os\label{ct}\\
&=& \sqrt\frac{\pi}{2\alpha}\sum_{i=1}^2\int \frac{dp}{2\pi}e^{-\frac{p^2}{4\alpha}}\sum_{j=1}^\infty\frac{(-it)^j}{j!} \omega_i(p)^{j-\frac{3}{2}}  |i,p\rangle\nonumber\\
&=& \sqrt\frac{\pi}{2\alpha}\sum_{i=1}^2\int \frac{dp}{2\pi}e^{-\frac{p^2}{4\alpha}}\frac{e^{-i\omega_i(p)t}-1}{\omega_i(p)^{\frac{3}{2}}}  |i,p\rangle\nonumber
\eea
where the $\supset$ on the first line is the restriction to terms with precisely one power of $H_I$.  We have chosen to omit terms with no powers of $H_I$ from (\ref{ct}) as they would not contribute to the matrix elements calculated below and, perhaps more to the point, they contain no neutrinos as so do not contribute to the neutrino wave packet.  According to the general arguments in Refs.~\cite{naumov1,naumov2}, one may identify the state $|t\rangle$ with a neutrino wave packet and expect that it is a covariant function of the four-momentum $p$.  No such covariance is manifest in Eq.~(\ref{ct}).  In a sequel, we will investigate whether the wave packets in our models possess the covariance property demanded in these references and assumed by the Daya Bay collaboration in their analysis \cite{dayadec}.

Note that we have not explicitly introduced the time $t_0$ when the neutrino is created.  However we may rewrite $|t\rangle$ as an integral over $t_0$
\beq
|t\rangle=-i\int_{t_0=0}^tdt_0\sqrt\frac{\pi}{2\alpha}\sum_{i=1}^2\int \frac{dp}{2\pi}\frac{e^{-\frac{p^2}{4\alpha}}}{\sqrt{\omega_i(p)}} e^{-i\omega_i(p)(t-t_0)} |i,p\rangle . \label{t0int}
\eeq

In this note we will not explicitly consider the measurements of neutrinos in our model, these will be included in future work.  Our goal for now is to understand neutrino wave functions.  These are already sufficient for constructing amplitudes and probabilities which will eventually be related to measurements in our companion paper.  We will be interested in the following amplitude, which corresponds to a transition to a neutrino at a position $x$ at time $t$
\bea
\mathcal{A}_i(x,t)&=&\langle i,x|t\rangle=\mo{q} e^{iqx} e^{-\frac{\left(q-p_0\right)^2}{2\sigma^2}} \sqrt\frac{\pi}{2\alpha}\sum_{j=1}^2\int \frac{dp}{2\pi}e^{-\frac{p^2}{4\alpha}}\left(\frac{e^{-i\omega_j(p)t}-1}{\omega_j(p)^{\frac{3}{2}}}\right)  \langle i,q|j,p\rangle\nonumber\\
&=&\sqrt\frac{\pi}{2\alpha}\mo{p}e^{ipx-\frac{\left(p-p_0\right)^2}{2\sigma^2} -\frac{p^2}{4\alpha}}\left(\frac{e^{-i\omega_i(p)t}-1}{\omega_i(p)^{\frac{3}{2}}}\right) . \label{caeq}
\eea
This amplitude, when $\sigma=\infty$, is the wave function of a single neutrino at time $t$.  

\begin{figure}
\begin{center}
\includegraphics[width=6.5in,height=.9in]{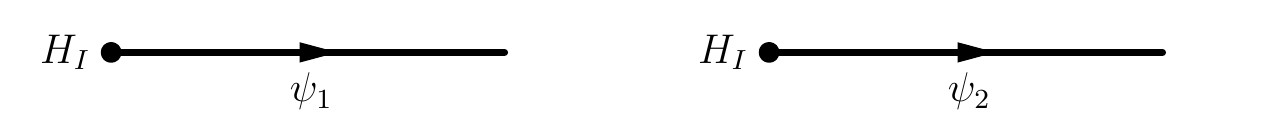}
\end{center}
\caption{The amplitudes $\mathcal{A}_1$ and $\mathcal{A}_2$ correspond to a neutrino which is created from the ground state of the free Hamiltonian $H_0$.}
\label{cafeynfig}
\end{figure}

Neutrinos are created in the flavor basis $\psi_1+\psi_2$.  Of course, it reality one also measures them in the flavor basis.  While we do not consider the measurement here, this does motivate us to introduce the flavor basis matrix element (shown in Fig.~\ref{cafeynfig})
\beq
\mathcal{A}(x,t)=\sum_{i=1}^2\mathcal{A}_i(x,t).
\eeq
One can also define a transition probability from the $H_0$ ground state to a one neutrino state.  This is not the probability of a measurement, since there is no term in our Hamiltonian which measures a neutrino.  It is simply the probability that a neutrino exists at time $t$ and position $x$, given that the system began in the $H_0$ ground state at time $t=0$.  Naively the transition probability would be
\beq
P(x,t)=\lambda |\mathcal{A}(x,t)|^2 \label{cpeq}
\eeq
where $\lambda$ is a normalization constant.  However since $x$ is continuous one expects that probability of finding a neutrino at any given $x$ to vanish, implying that $\lambda=0$.  For a continuous $x$ one is interested instead in the probability density $dP(x,t)/dx$. 
\unitlength = 1mm

It is therefore tempting to identify (\ref{cpeq}) with a probability density where
\beq
\lambda=\frac{2}{\sigma^2} \label{lambda}
\eeq
to cancel the normalization in (\ref{prodeq}).  This is not quite right, due to the fact that the $|i,x\rangle$ basis is not orthogonal and so if the neutrino is observed at $x$ it has a nonzero probability to also be observed at $y$.  Therefore one cannot define a normalized probability distribution function (PDF) for $x$.  However, such double-valued position probabilities are exponentially suppressed at distances larger than the de Broglie wavelength corresponding to the momentum resolution.  The position resolution of any neutrino detector is much larger than this distance, and so for all practical purposes (\ref{cpeq}) is a PDF.   

Summarizing, we have argued that the transition probability density for the creation of a neutrino at $(x,t)$ in the flavor basis is
\beq
P(x,t)=\frac{\pi}{\alpha\sigma^2}\left|\sum_{i=1}^2\mo{p}e^{ipx-\frac{\left(p-p_0\right)^2}{2\sigma^2} -\frac{p^2}{4\alpha}}\left(\frac{e^{-i\omega_i(p)t}-1}{\omega_i(p)^{\frac{3}{2}}}\right) \right|^2 . \label{cpeq2}
\eeq
This can again be written as an integral over the interaction time $t_0$
\beq
P(x,t)=\frac{\pi}{\alpha\sigma^2}\left|\int_{t_0=0}^t dt_0\sum_{i=1}^2\mo{p}\frac{e^{ipx-\frac{\left(p-p_0\right)^2}{2\sigma^2} -\frac{p^2}{4\alpha}}}{\sqrt{\omega_i(p)}}e^{-i\omega_i(p)(t-t_0)} \right|^2. 
\eeq
The largest contribution to the integral over $p$ comes from the stationary point of the phase 
\beq
\frac{\partial}{\partial p}(px-\omega_i(p)(t-t_0))=0
\eeq
and so
\beq
\frac{\partial}{\partial p}\omega_i(p)=\frac{x}{t-t_0}
\eeq
which yields the usual condition that the group velocity is equal to the average velocity of the neutrino in the time $t-t_0$ since its creation.

\subsection{Numerical Results}

We will now consider the case
\beq
\alpha=1\hsp p_0=1\hsp \sigma=0.3\hsp m_1=0.3\hsp m_2=0.4
\eeq
corresponding to classical source of width $0.5$, a measured neutrino momentum of $1\pm 0.1$ and neutrino masses of $0.3$ and $0.4$.  

\begin{figure} 
\begin{center}
\includegraphics[width=2.5in,height=1.7in]{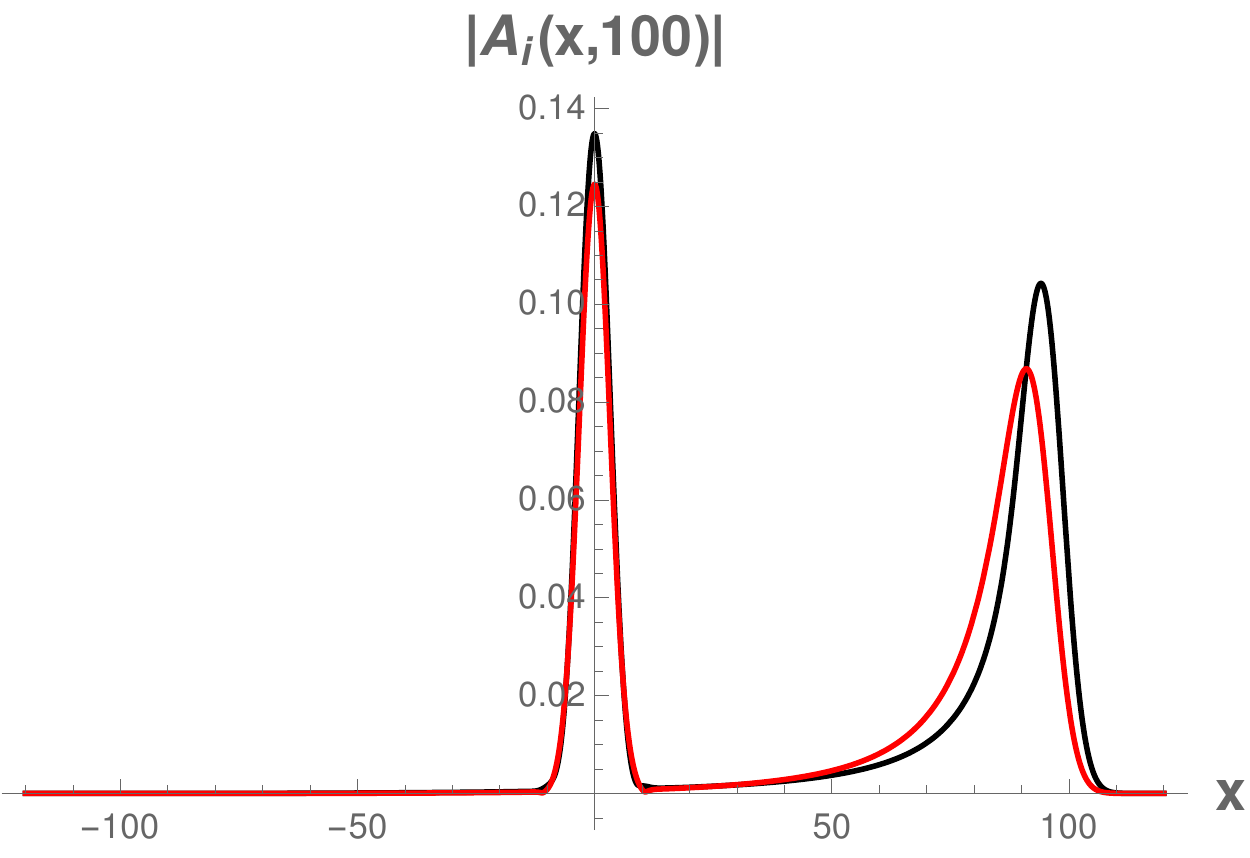}
\includegraphics[width=2.5in,height=1.7in]{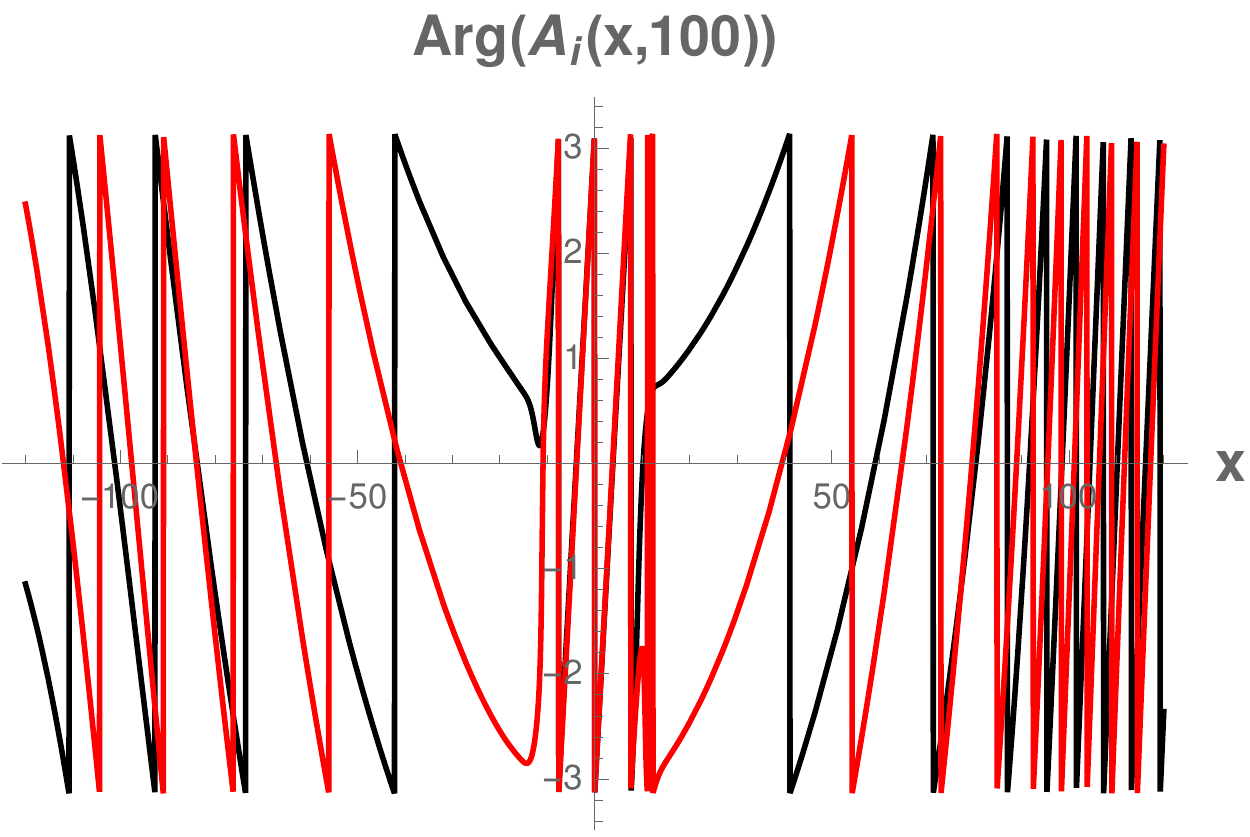}
\caption{The absolute values (left) and phases (right) of the amplitudes $A_i(x,100)$ calculated at time $t=100$ in the classical source model.  The black and red curves correspond to the $m=0.3$ and $m=0.4$ neutrino wave functions respectively.}
\label{cafig}
\end{center}
\end{figure}

The amplitudes $\mathcal{A}_i(x,100)$ defined in Eq.~(\ref{caeq}) are shown in Fig.~\ref{cafig}.  Three peculiar features are evident in the left panel.  First, the maximum amplitude occurs near $x=0$.  This is a consequence of the fact that the initial energy of the system is equal to zero, since $H_0$ annihilates the initial state $\Os$.  The final energy is therefore also equal to zero, as $H$ is time-independent and so time evolution conserves energy.  However the neutrinos are massive, and so they will always be off-shell.  This is reflected in the $\omega$ in the denominator, which vanishes only if $\omega=0$, as is never the case.  The smallest $\omega$ is the least off-shell, and therefore the highest amplitude.   As a result the highest amplitude arises for the neutrinos with the smallest momentum, which cannot travel far.

The second and least physical peculiar feature is the peak near $x=t$ corresponding to neutrinos created at $t_0=0$.  Recall from Eq.~(\ref{t0int}) that one integrates over $t_0$, and so why should most of the neutrinos observed arise from $t_0\sim 0$?  This is another consequence of the fact that the neutrinos are off-shell.  As $\omega\neq 0$, the phase $e^{-i\omega t}$ in Eq.~(\ref{t0int}) always oscillates, damping the integral and so the amplitude.  This damping is reduced at $t_0=0$ just because this is a boundary of the domain of integration, and so there is no oscillation at $t_0<0$.  

The third peculiar feature is the small tail at $x>t$.  One may attribute this tail to the finite size $1/(2\sqrt\alpha)$ of the classical source.  However the tail is too large to be created by this alone.  It is also a consequence of the fact that $\mathcal{A}$ is essentially the Feynman propagator $\langle\Omega|\psi(t)\psi(t_0)\Os$, albeit with some additional factors.  Recall that in quantum field theory only the retarded propagator is causal.  The causality of the retarded propagator results from the presence of a commutator term $-\langle\Omega|\psi(t_0)\psi(t)\Os$.  However no such term is present in $\mathcal{A}$.  The physical explanation for the lack of causality of the Feynman propagator is that a particle of mass $m$ cannot be kept in a box of size beneath $1/m$, and so a leaking of order $1/m$ is inevitable \cite{colemanlect}.  Despite the small mass of the neutrino, the length scale $1/m$ is far smaller than the position resolution of any experiment and so this tail is irrelevant in neutrino physics.

\begin{figure} 
\begin{center}
\includegraphics[width=2.5in,height=1.7in]{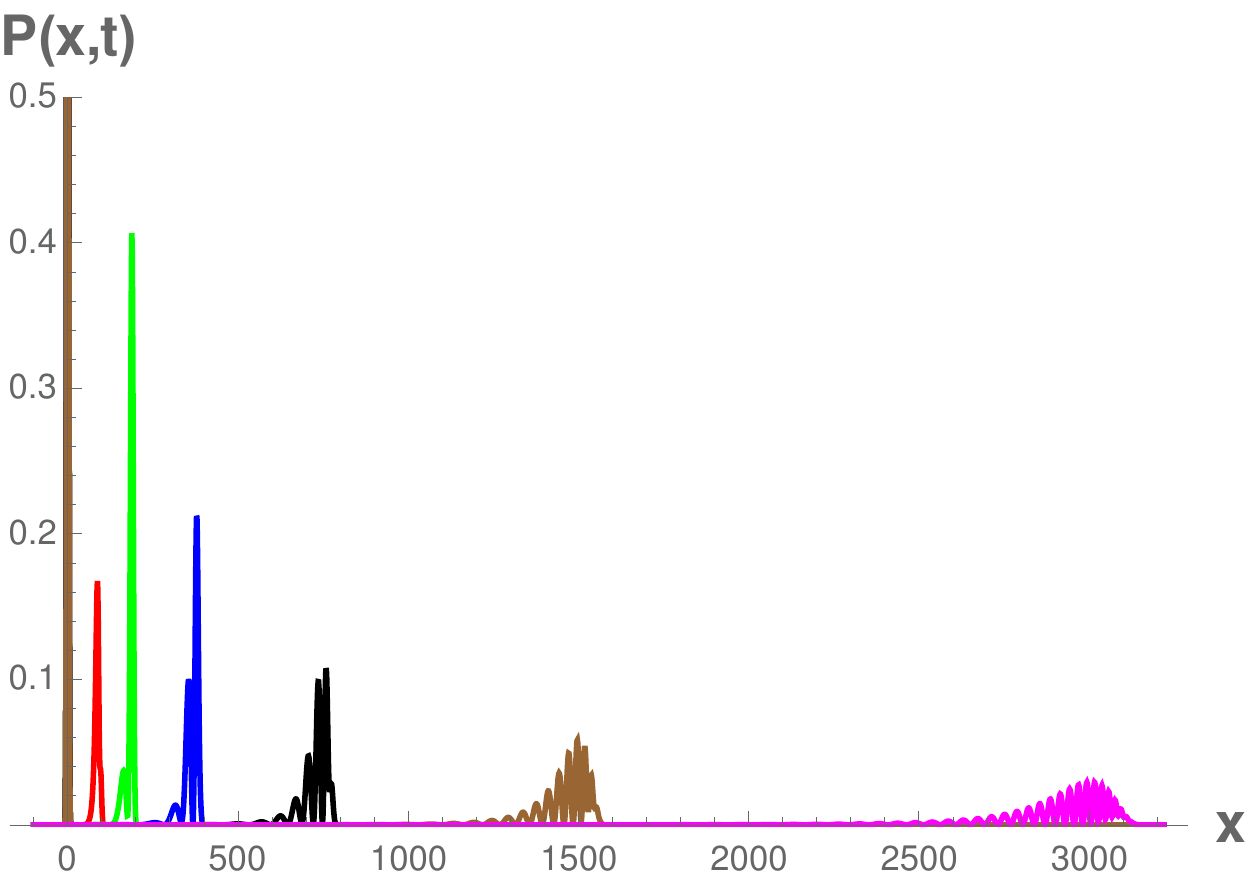}
\caption{The probability density $P(x,t)$ at time $t=100,\ 200,\ 400,\ 800,\ 1600$ and $3200$ in red, green, blue, black, brown and magenta respectively.  As expected, neutrinos oscillate more at later times.}
\label{cptuttifig}
\end{center}
\end{figure}

\begin{figure} 
\begin{center}
\includegraphics[width=2.5in,height=1.7in]{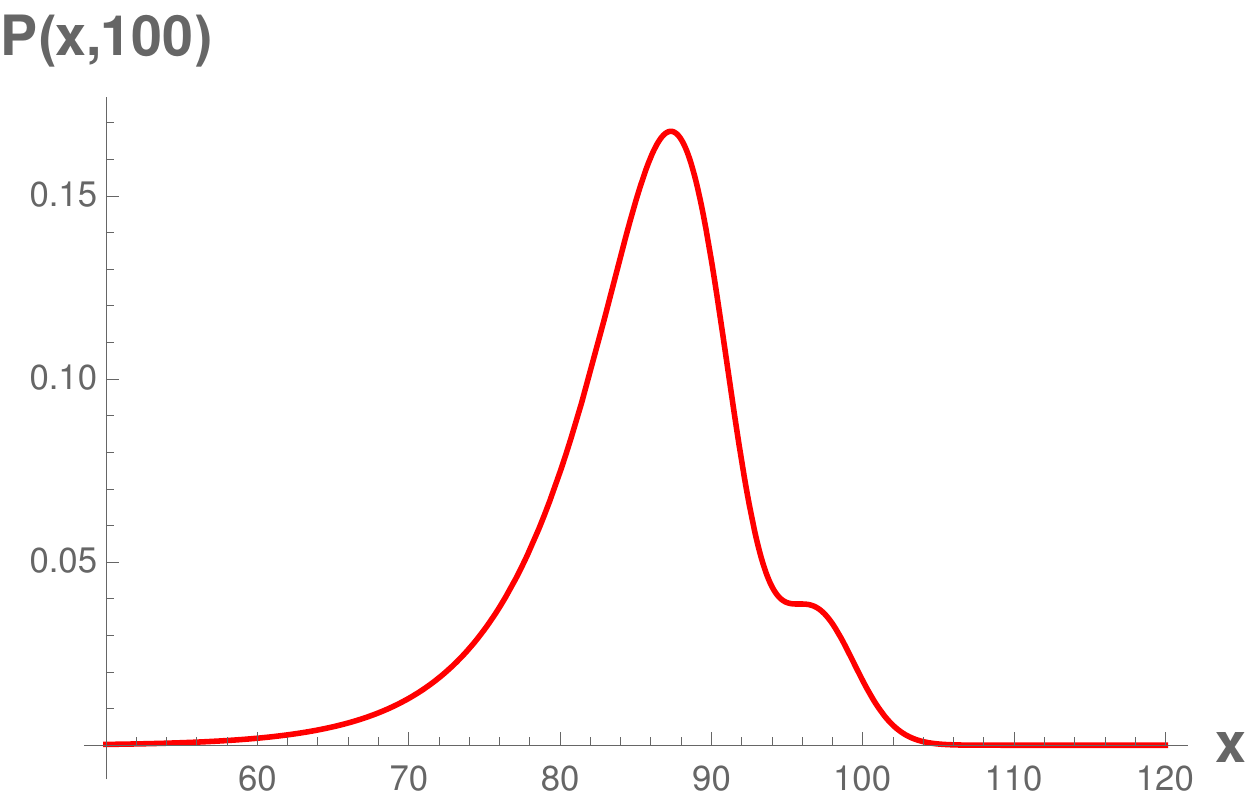}
\includegraphics[width=2.5in,height=1.7in]{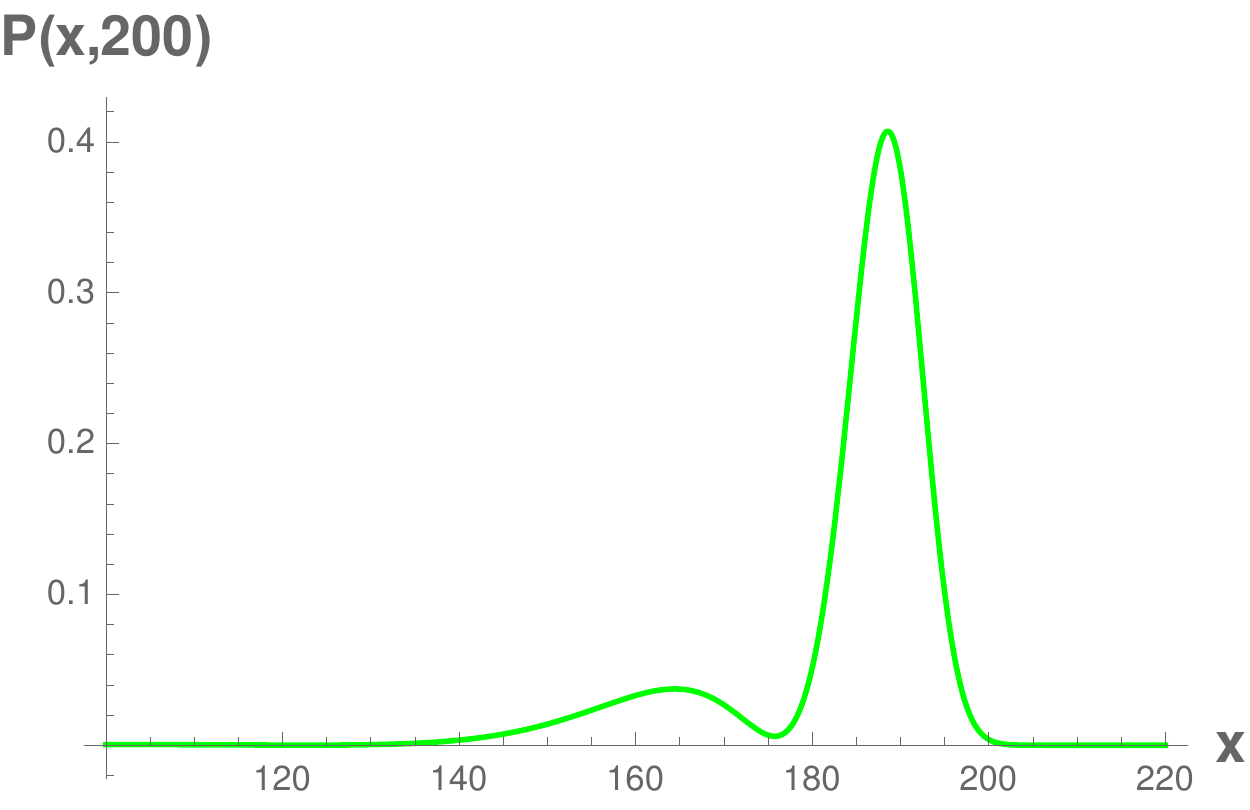}
\includegraphics[width=2.5in,height=1.7in]{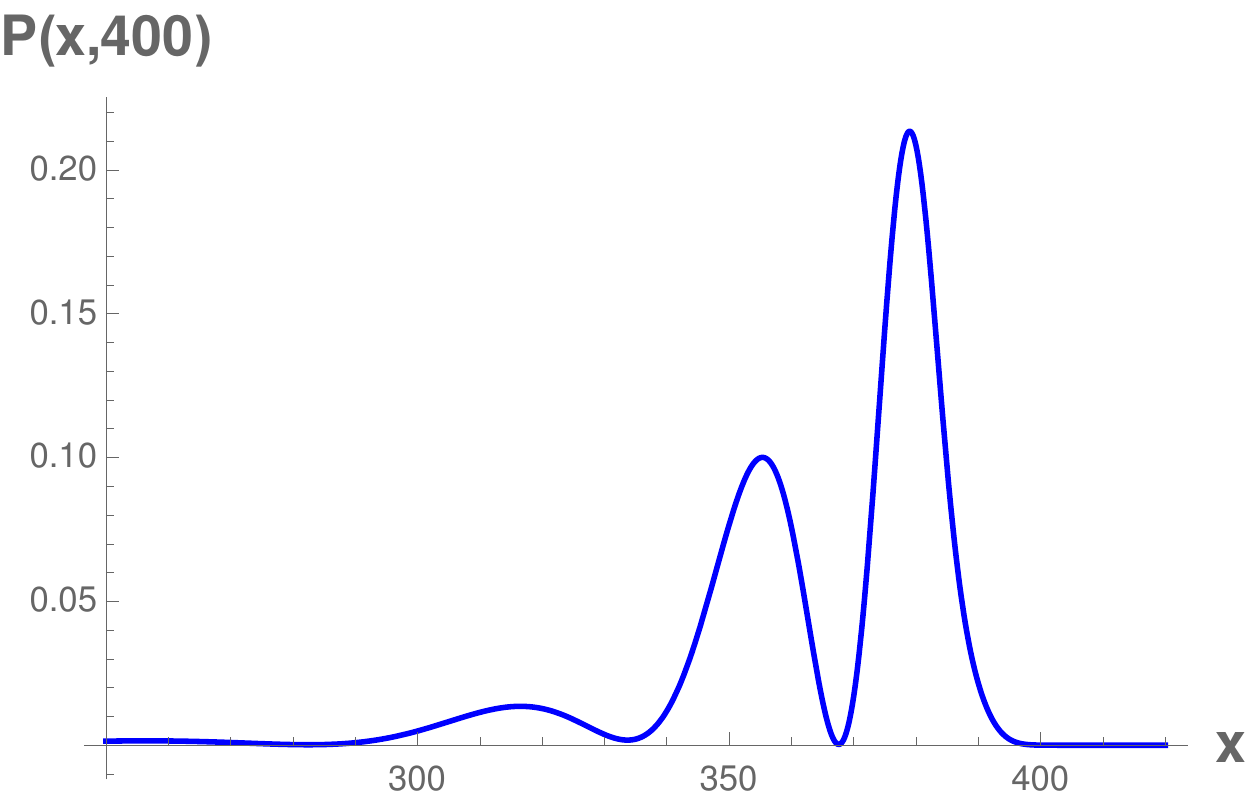}
\includegraphics[width=2.5in,height=1.7in]{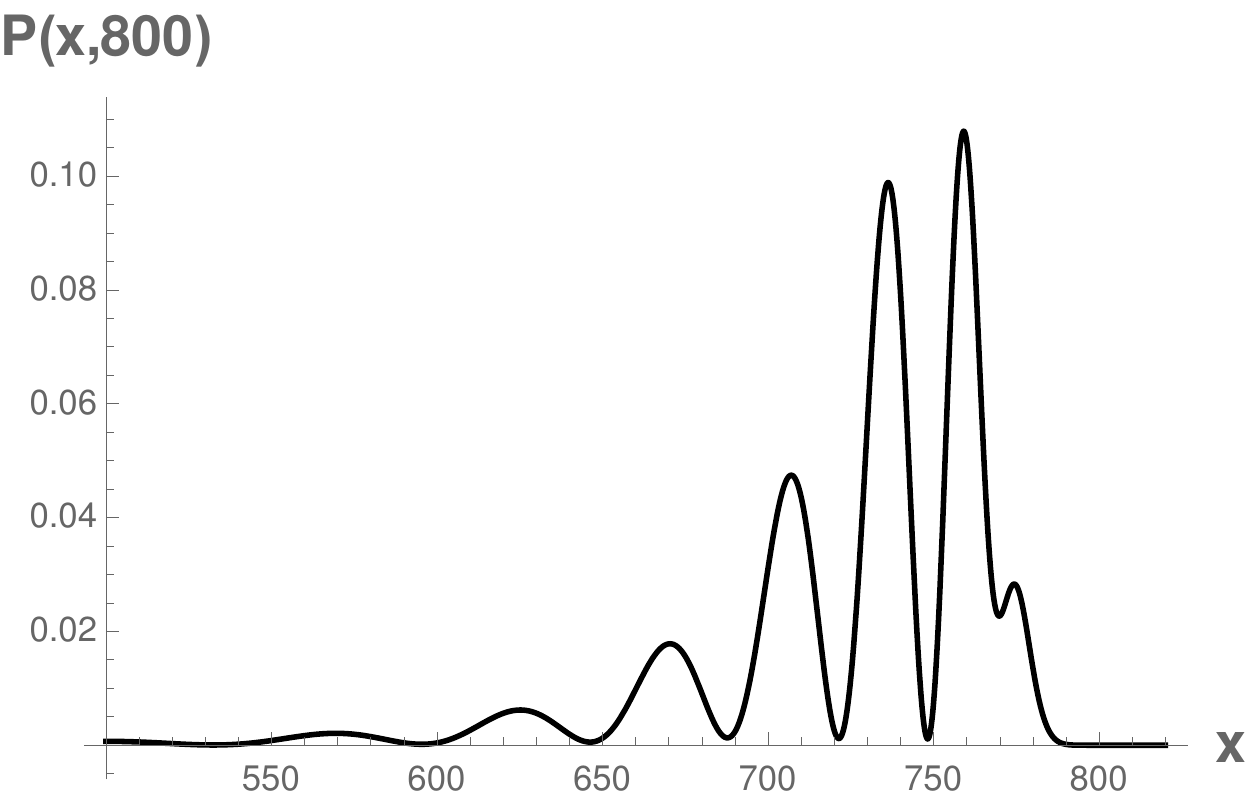}
\includegraphics[width=2.5in,height=1.7in]{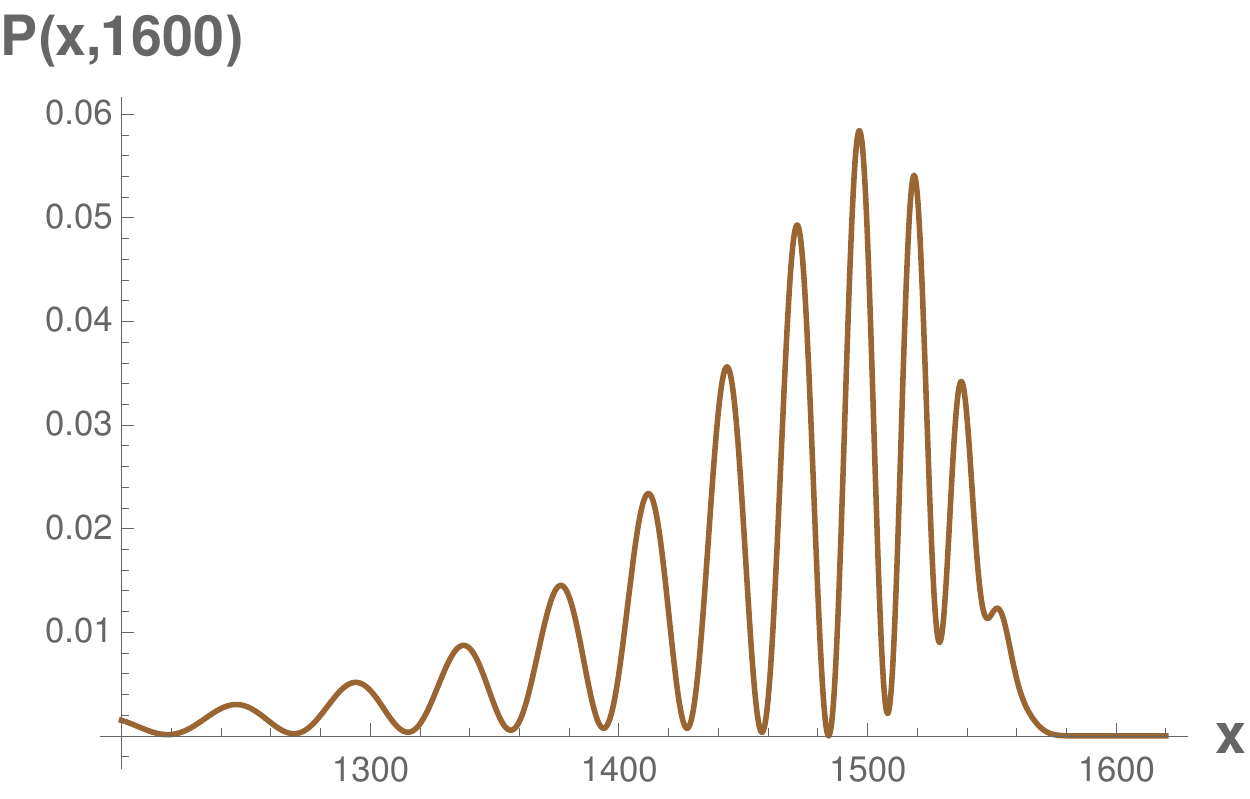}
\includegraphics[width=2.5in,height=1.7in]{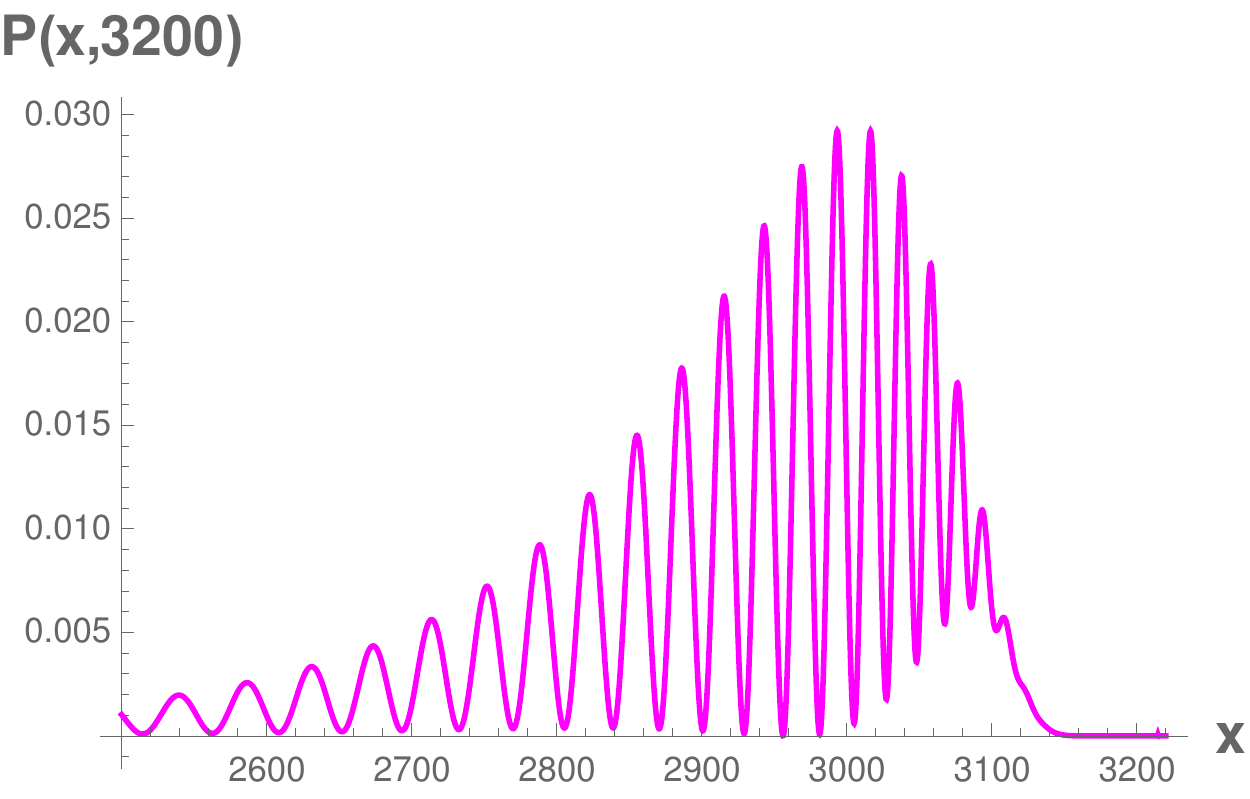}
\caption{As in Fig.~\ref{cptuttifig} but each time is shown in its own panel.}
\label{cpfig}
\end{center}
\end{figure}

The resulting probability densities, as summarized in Eq.~(\ref{cpeq2}), are shown in Figs.~\ref{cptuttifig} and \ref{cpfig}.  The three peculiar features seen in the amplitudes are also present in the probabilities.  However, neutrino oscillations are clearly present and, as expected, are more numerous at late times.   The slight damping of the oscillations near the light cone results from the fact, already visible in Fig.~\ref{cafig}, that the more massive neutrino travels more slowly and so its amplitude is smaller than that of the lighter neutrino near the light cone. Such kinematic damping is far too small to observe at present day neutrino experiments.

\section{The Model} \label{modsez}

We are interested in decoherence resulting from interactions of the source particle with the environment, together with quantum entanglement between the neutrino, the source and the environment.  The source above was classical and so could not be entangled.  Therefore, to incorporate decoherence in our model we must introduce quantum source fields $\phi_I$ and environment fields $E_\alpha$.

\subsection{The Fields and Their Interactions}

In oscillation experiments the neutrinos travel macroscopic distances and so are observed on-shell.  While we do not assert that our final states are on-shell, it will be clear from our expressions that off-shell final states will generally provide a small contribution.  The simplest on-shell decay in a Lorentz-invariant theory is the decay of a heavy source particle $\phi_H$ into a slightly lighter yet still heavy particle $\phi_L$ and our so-called neutrino $\psi_i$, which is actually a scalar.  Neutrino oscillations require at least two values of the index $i$ which labels neutrino mass eigenstates.  Thus the simplest model with oscillations contains four real scalar fields $\phi_H$, $\phi_L$, $\psi_1$ and $\psi_2$ with masses $M_H>M_L>m_i$ together with the interaction Hamiltonian
\beq
\mathcal{H}_I(x)=\phi_H(x)\phi_L(x)\left(\psi_1(x)+\psi_2(x)\right). \label{hi}
\eeq
Unlike real-world $\beta$ decay, the neutrinos in our model are created in a two-body process in which $\phi_H$ decays to $\phi_L$ and $\psi_i$.

Decoherence requires coupling to environment fields $E_\alpha$.  While two fields would be sufficient, we will consider four, indexed by $\alpha\in[0,3]$.  These will interact with $\phi_H$ via interactions of the form $\epsilon_\alpha \phi^2_H E_\alpha^2$.  We will consider a nonrelativistic approximation of this interaction, so that it is of the form of that in Ref.~\cite{zurek}.  In this approximation, we simply add a perturbation to the Hamiltonian equal to 
\beq
H^\prime=\sum_{\alpha} \epsilon_\alpha N_H N_\alpha
\eeq
where $N_H$ and $N_\alpha$ are the usual particle number operators for the fields $\psi_H$ and $E_\alpha$.

\subsection{The States}

We will perform the usual decomposition of the canonical fields
\bea
\phi_I(x)&=&\int\frac{dp}{2\pi}\frac{1}{\sqrt{2\Omega_I}}\left(A_{I,-p}+A^\dagger_{I,p}\right)e^{-ipx}\hsp \Omega_I(p)=\sqrt{M_I^2+p^2}\nonumber\\
\Pi_I(x)&=&-i\int\frac{dp}{2\pi}{\sqrt{\frac{\Omega_I}{2}}}\left(A_{I,-p}-A^\dagger_{I,p}\right)e^{-ipx}
\eea
where $I$ runs over the indices $\{H,L\}$.  The decomposition of the environment fields will not be needed due to our nonrelativistic approximation. 

We will only be interested in states with one environmental particle $E_\alpha$, one source particle $\psi_I$ and zero or one neutrinos $\phi_i$.  We will not keep track of the momentum or the position of the environmental particle, we will only be interested in its flavor $\alpha$.  Thus a basis of the states of interest may be written $|\alpha;I,p;i,q\rangle$ for states with a neutrino of flavor $i$ and momentum $q$ and a source particle of flavor $I$ and momentum $p$, together with the states $|\alpha;I,p\rangle$ which contain no neutrino.   The free particle ground states, with an environment field, may be written as simply $|\alpha\rangle$.  These are annihilated by all operators $a$ and $A$ and are orthonormal.  The normalizations of the other states are fixed by
\beq
|\alpha;I,p\rangle=A^\dagger_{I,p}|\alpha\rangle\hsp
|\alpha;I,p;i,q\rangle=A^\dagger_{I,p}a^\dagger_{i,q}|\alpha\rangle .
\eeq

Our initial condition will consist of a heavy source particle in a Gaussian wave packet
\beq
|0\rangle=\sum_\alpha c_\alpha \int\frac{dp}{2\pi}e^{-p^2/(4\beta)}|\alpha;H,p\rangle
\eeq
where $\beta$ is a parameter which determines the initial width of the wave packet.  This state is normalized such that
\beq
\langle 0|0\rangle=\sqrt\frac{\beta}{2\pi}\sum_\alpha c_\alpha^2 .
\eeq
One could fix the $c_\alpha$ so that this is equal to unity, but we will instead leave the $c_\alpha$ free and correct for this normalization in our formula for the probability.

The initial state $|0\rangle$ will evolve into states $|\alpha;L,p;i,q\rangle$ and so we will be interested in matrix elements of the form $\langle\alpha;L,p;i,q|e^{-iHt}|0\rangle$ where $H$ is the total Hamiltonian and $t$ is the time to which the system evolves.   This matrix element is the amplitude, calculated in the Schrodinger picture, for the initial state $|0\rangle$ to evolve into the final state $|\alpha;L,p;i,q\rangle$.

\begin{figure}
\begin{center}
\includegraphics[width=6.5in,height=1.5in]{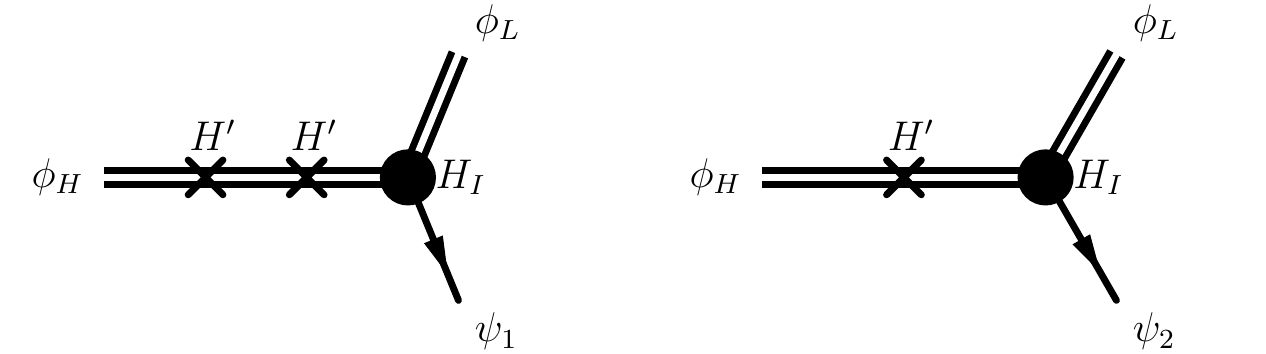}
\caption{The amplitudes $\mathcal{A}_{1\alpha}$ (left) and $\mathcal{A}_{2\alpha}$ (right) correspond to a neutrino $\psi_1$ and $\psi_2$ respectively which is created from a single source particle $\phi_H$ interacting with an environment in the state $|\alpha\rangle$.   If $\psi_2$ is heavier than $\psi_1$, it will be slower at fixed momentum and so, given an observation at a fixed time and position, it is emitted earlier \cite{mcgreevy}.  Therefore $\phi_H$ has less time to interact with the environment in $\mathcal{A}_{2\alpha}$ than in $\mathcal{A}_{1\alpha}$.}
\label{pfeynfig}
\end{center}
\end{figure}

In this paper we will not yet introduce neutrino measurements.  However, as our interest does nonetheless lie in measurement, we consider matrix elements which are close to the measured quantities.  Since any measurement will be in the flavor basis and not the mass basis, we will sum these matrix elements over $i$ corresponding to a disappearance channel experiment.  Also, since a measurement will also measure, with some resolution $\sigma$, the neutrino momentum $q$ and will find a value $q_0$, we will calculate the matrix elements (shown in Fig.~\ref{pfeynfig})
\beq
\mathcal{A}_\alpha(p,x,t)=\sum_{i=1}^2\mathcal{A}_{i\alpha}(p,x,t)\hsp
\mathcal{A}_{i\alpha}(p,x,t)=\mo{q} e^{-(q-q_0)^2/(2\sigma^2)}e^{iqx}\langle\alpha;L,p;i,q|e^{-iHt}|0\rangle . \label{adef}
\eeq
If $\sigma\neq \infty$ then these will not be precisely orthogonal.  However, we will ignore this and define an approximate probability density as
\beq
P(x,t)=\sum_\alpha P_\alpha(x,t)\hsp
P_\alpha(x,t)=\lambda\int \frac{dp}{2\pi} | \mathcal{A}_\alpha(x,p)|^2 \label{peq}
\eeq
where $\lambda$ is the normalization constant
\beq
\lambda=\frac{\sqrt{2}}{\sigma\sqrt{\beta}}\frac{1}{\sum_\alpha |c_\alpha^2|}.
\eeq
Note that, unlike the wave packet approach \cite{review,giunti2012} in which one considers only a single final state for the source particle, here the final state $p$ is integrated over.  This is reasonable as the final state of the source particle is never measured, and as a result the neutrino is never in a localized wave packet.

These amplitudes and probability densities correspond to transitions from the heavy particle to the light particle plus a neutrino.  These are the usual transition amplitudes and transition probabilities in quantum field theory.   These are not equal to the amplitudes or probabilities for neutrino measurement, which would require an additional interaction in which the neutrino is absorbed.  Nonetheless, these amplitudes and probability densities are interesting because they already manifest neutrino oscillations and decoherence and therefore provide a simple setting in which these pheneomena may be studied.

\section{Results} \label{ressez}

\subsection{Analytical Calculation}

As events involving multiple neutrinos are suppressed by the Fermi coupling constant, we will work only to linear order in $H_I$ and will consider only 0-neutrino and 1-neutrino states.  Therefore it will be convenient to decompose the Hamiltonian into a neutrino-number conserving piece $H_0$ and the neutrino creating term $H_I$ given in Eq.~(\ref{hi})
\beq
H=H_0+H_I\hsp H_0=H^\prime+\int dx \mathcal{H}_0(x)
\eeq
where $\mathcal{H}_0(x)$ is the free real scalar Hamiltonian density
\beq
\mathcal{H}_0(x)=\frac{1}{2}\sum_{i=1}^2:\left(\pi_i(x)^2+\left(\partial_x\psi(x)\right)^2+m_i^2\psi_i^2\right):
+\frac{1}{2}\sum_{I=H,L}:\left(\Pi_I(x)^2+\left(\partial_x\phi_I(x)\right)^2+M_I^2\phi_I^2\right):.
\eeq
The neutrino-number conserving Hamiltonian is then
\beq
H_0=\mo{p}\left[\sum_{i=1}^2\omega_i(p)a^\dagger_{i,p}a_{i,p}+\sum_{\alpha=0}^3 \epsilon_\alpha N_\alpha A^\dagger_{H,p}A_{H,p}+\sum_{I=H,L}\Omega_I(p) A^\dagger_{I,p}A_{I,p}\right].
\eeq

Our $0$ and $1$-neutrino basis of states are again eigenstates of $H_0$
\beq
H_0|\alpha;H,p\rangle=E_{0,\alpha}(p)|\alpha;H,p\rangle\hsp
H_0|\alpha;L,p;i,q\rangle=E_{1,i}(p,q)|\alpha;L,p;i,q\rangle
\eeq
where we have defined the eigenvalues
\beq
E_{0,\alpha}(p)=\Omega_H(p)+\epsilon_\alpha\hsp
E_{1,i}(p,q)=\Omega_H(p)+\omega_i(q).
\eeq
The interaction $H_I$ interpolates between these two sectors
\beq
H_I|\alpha;H,p\rangle=\sum_{i=1}^2\mo{q}\frac{|\alpha;L,q;i,p-q\rangle}{\sqrt{8\Omega_{H}(p)\Omega_L(q)\omega_i(p-q)}}.
\eeq

The evolution of a $0$-neutrino state is slightly more complicated than in the classical source case because $H_0$ does not annihilate the initial configuration, which now contains both a source particle and also an environment particle.  Again, restricting attention to terms with precisely one $H_I$ we find
\bea
e^{-iHt}|\alpha;H,p\rangle&=&\sum_{k=0}^\infty \frac{(-iHt)^k}{k!}|\alpha;H,p\rangle
\supset\sum_{k=1}^\infty\sum_{j=0}^{k-1} \frac{(-it)^k}{k!}H_0^{j}H_IH_0^{k-j-1}|\alpha;H,p\rangle\label{t}\\
&=&\sum_{k=1}^\infty\frac{(-it)^k}{k!}\sum_{j=0}^{k-1}E_{0,\alpha}(p)^{k-j-1} H_0^{j}H_I|\alpha;H,p\rangle\nonumber\\
&=&\sum_{i=1}^2\mo{q}\sum_{k=1}^\infty\frac{(-it)^k}{k!}\sum_{j=0}^{k-1}E_{0,\alpha}(p)^{k-j-1}H_0^{j}\frac{|\alpha;L,q;i,p-q\rangle}{\sqrt{8\Omega_{H}(p)\Omega_L(q)\omega_i(p-q)}}\nonumber\\
&=&\sum_{i=1}^2\mo{q}\left(\sum_{k=1}^\infty\frac{(-it)^k}{k!}\sum_{j=0}^{k-1}E_{0,\alpha}(p)^{k-j-1}E_{1,i}(q,p-q)^{j}\right)\frac{|\alpha;L,q;i,p-q\rangle}{\sqrt{8\Omega_{H}(p)\Omega_L(q)\omega_i(p-q)}}\nonumber\\
&=&\sum_{i=1}^2\mo{q}\left(\frac{e^{-iE_{1,i}(q,p-q)t}-e^{-iE_{0,\alpha}(p) t}}{E_{1,i}(q,p-q)-E_{0,\alpha}(p)}\right)\frac{|\alpha;L,q;i,p-q\rangle}{\sqrt{8\Omega_{H}(p)\Omega_L(q)\omega_i(p-q)}} .\nonumber
\eea
Again we remind the reader that the terms with no $H_0$ have been dropped as they will not contribute to the matrix elements calculated below and also, as they contain no neutrinos, they will not contribute to our understanding of the neutrino wave packet.  The projected state (\ref{t}) can be written in terms of an integral over the time $t_0$ at which the neutrino was created
\beq
e^{-iHt}|\alpha;H,p\rangle=-i\sum_{i=1}^2\mo{q}\frac{e^{-iE_{1,i}(q,p-q) t}|\alpha;L,q;i,p-q\rangle
}{\sqrt{8\Omega_{H}(p)\Omega_L(q)\omega_i(p-q)}}\int_{t_0=0}^tdt_0e^{-i\left(E_{0,\alpha}(p)-E_{1,i}(q,p-q)\right) (t-t_0)}.\label{t0}
\eeq
A measurement of a neutrino at a specific $(x,t)$ would allow a determination of $t_0$ to within some uncertainty.  However no measurement is implied here and so all values of $t_0\in[0,t]$ contribute to the amplitudes.

The Hamiltonian is again time-independent and so evolution conserves energy.  $E_0$ and $E_1$ are not precisely the energies of the initial and final states, but rather the energies that they would have were they on-shell.  The phase in (\ref{t0}) oscillates rapidly in $t_0$ unless $E_0=E_1$.  Therefore the $t_0$ integral will be dominated by the stationary phase corresponding to the case in which the particles are on-shell.  In this way we naturally recover the fact that particles are on-shell when $t$ is large.  This is also apparent in Eq.~(\ref{t}), where the $(E_1-E_0)$ in the denominator favors $E_0\sim E_1$.  Note that there is no pole as the numerator vanishes when $E_0=E_1$.

As the evolution operator $e^{-iHt}$ is linear, one can now easily evaluate the state at a time $t$
\bea
|t\rangle &=&e^{-iHt}|0\rangle\\
&=&\sum_\alpha c_\alpha \int\frac{dp}{2\pi}e^{-p^2/(4\beta)}\sum_{i=1}^2\mo{q}\left(\frac{e^{-iE_{1,i}(q,p-q)t}-e^{-iE_{0,\alpha}(p) t}}{E_{1,i}(q,p-q)-E_{0,\alpha}(p)}\right)\frac{|\alpha;L,q;i,p-q\rangle}{\sqrt{8\Omega_{H}(p)\Omega_L(q)\omega_i(p-q)}} .\nonumber
\eea
The 3-momentum of the neutrino is $p-q$.  The covariant wave packet conjecture states that $|t\rangle$ only depends on $p-q$ via Lorentz scalars.  This is certainly not evident, but we will test this claim in the sequel, beginning with an initial condition which is itself a covariant wave packet.

Again we calculate the matrix elements corresponding to transitions to states with neutrinos in the flavor basis.  The momentum space matrix elements are
\beq
\tilde{\mathcal{A}}_{i\alpha}(p,q,t)=\langle \alpha;L,p;i,q|t\rangle
=c_\alpha \frac{e^{-(p+q)^2/(4\beta)}}{\sqrt{8\Omega_{H}(p+q)\Omega_L(p)\omega_i(q)}}\left(\frac{e^{-iE_{1,i}(p,q)t}-e^{-iE_{0,\alpha}(p+q) t}}{E_{1,i}(p,q)-E_{0,\alpha}(p+q)}\right)
\eeq
where $p$ is the final momentum of the source and $q$ is the momentum of the neutrino.  In neutrino measurements, often both the position and the momentum of the neutrino are determined with some known uncertainty.  This motivates us to consider a transition amplitude in which both the momentum and the position of the neutrino are fixed, as in Eq.~(\ref{adef})
\beq
\mathcal{A}_{i\alpha}(p,x,t)
=c_\alpha \mo{q} e^{-(q-q_0)^2/(2\sigma^2)}e^{iqx}\frac{e^{-(p+q)^2/(4\beta)}}{\sqrt{8\Omega_{H}(p+q)\Omega_L(p)\omega_i(q)}}\left(\frac{e^{-iE_{1,i}(p,q)t}-e^{-iE_{0,\alpha}(p+q) t}}{E_{1,i}(p,q)-E_{0,\alpha}(p+q)}\right).
\eeq
Eq.~(\ref{peq}) then yields the approximate probability density for a transition to a state with a neutrino at position $x$ with momentum $q_0$
\bea
P(x,t)&=&\frac{\sqrt{2}}{\sigma\sqrt{\beta}}\frac{1}{\sum_\beta |c_\beta^2|}\sum_\alpha |c_\alpha^2|\mo{p}\\
&\times& \left|\sum_i \mo{q} e^{-(q-q_0)^2/(2\sigma^2)}e^{iqx}\frac{e^{-(p+q)^2/(4\beta)}}{\sqrt{8\Omega_{H}(p+q)\Omega_L(p)\omega_i(q)}}\left(\frac{e^{-iE_{1,i}(p,q)t}-e^{-iE_{0,\alpha}(p+q) t}}{E_{1,i}(p,q)-E_{0,\alpha}(p+q)}\right)\right|^2.\nonumber
\eea
We repeat that this is not the probability that the neutrino is measured, as no neutrinos are measured in our model.

In our model the environment only interacts with $\phi_H$, producing a shift in $E_{0,\alpha}(p+q)$ and so a relative phase between the two terms in the numerator on the right.  This is ultimately responsible for decoherence.  If, on the other hand, we introduce an additional coupling of the environment to both $\phi_H$ and $\phi_L$ with equal coefficients, it would produce an equal shift in both $E_{1,i}(p,q)$ and $E_{0,\alpha}(p+q)$.  The result would be an overall phase in the amplitude, which of course does not affect $P(x,t)$ as this only depends on the absolute value of the amplitude.  Therefore in this simple model we see that it is not the total interaction of the source with the environment which contributes to decoherence, as has been assumed in many calculations of decoherence such as Refs.~\cite{nuss76,wilczek}, but rather the difference between the interaction with the source state before and after the neutrino production.  In the case of a Coulomb interaction with a nucleus that produces a neutrino via $\beta$ decay, this would correspond to the difference in the Coulomb interaction caused by a shift in the charge $Z$ by one unit and the creation of a positron.  We claim that this factorization argument is quite general, and not a specific feature of our model. 

\subsection{Numerical Results: Amplitudes}

\begin{figure} 
\begin{center}
\includegraphics[width=2.5in,height=1.7in]{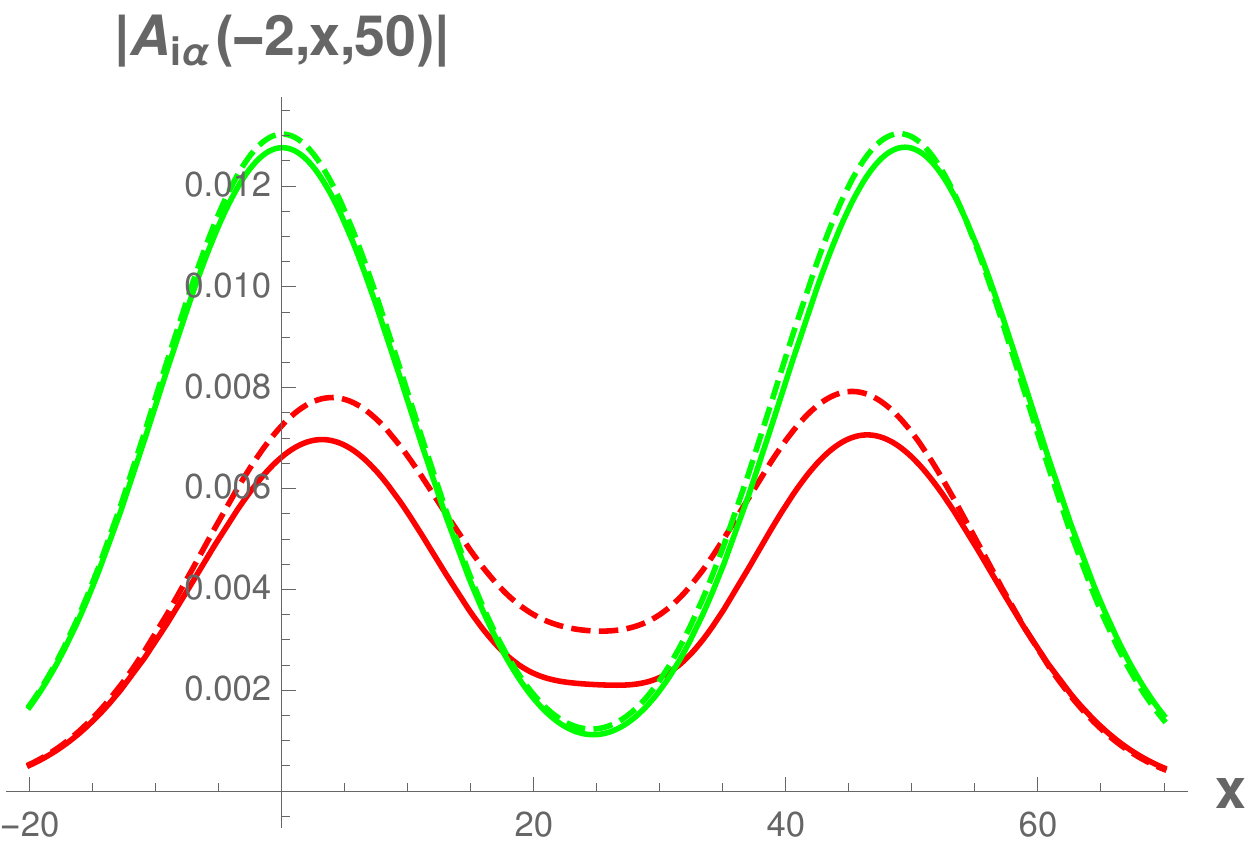}
\includegraphics[width=2.5in,height=1.7in]{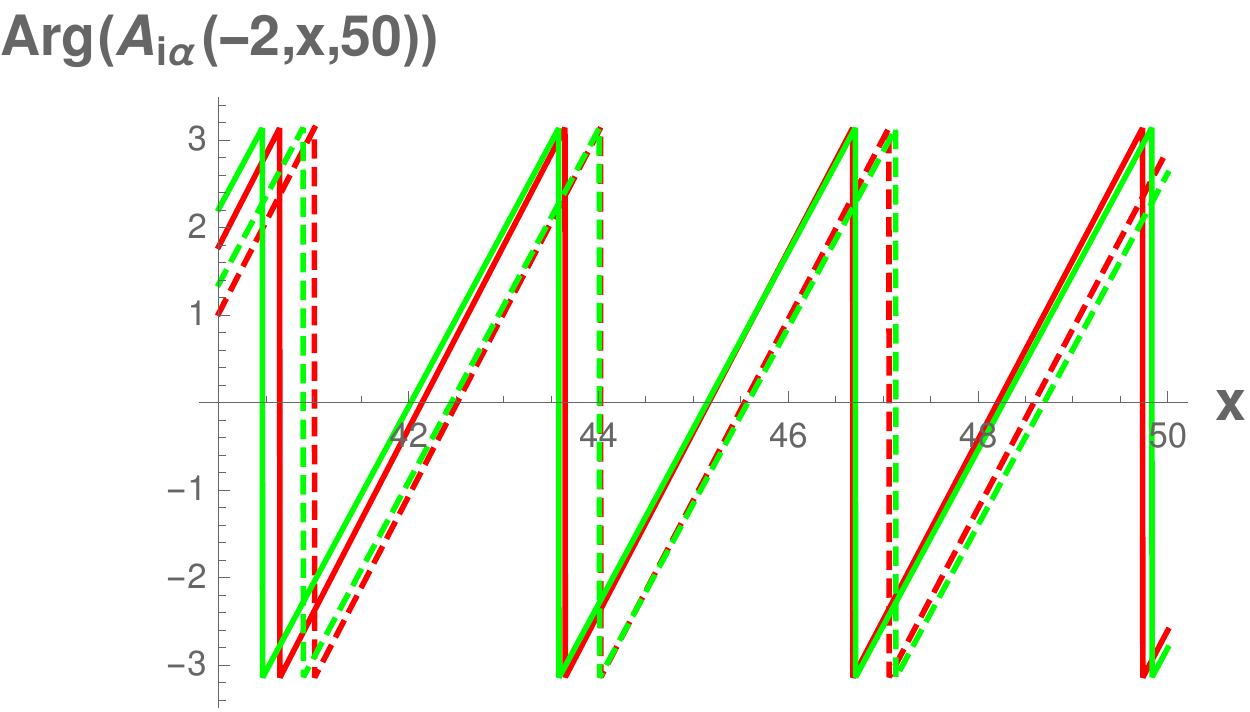}
\includegraphics[width=2.5in,height=1.7in]{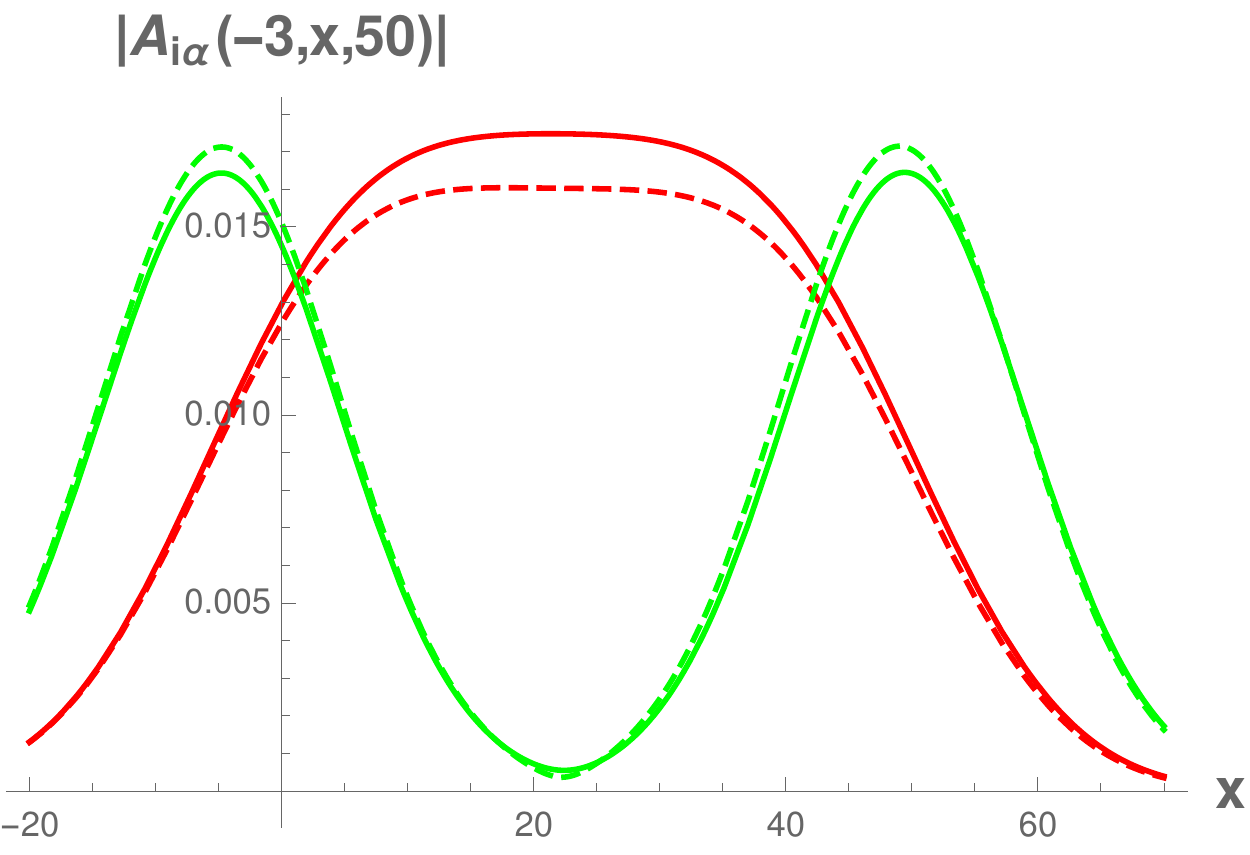}
\includegraphics[width=2.5in,height=1.7in]{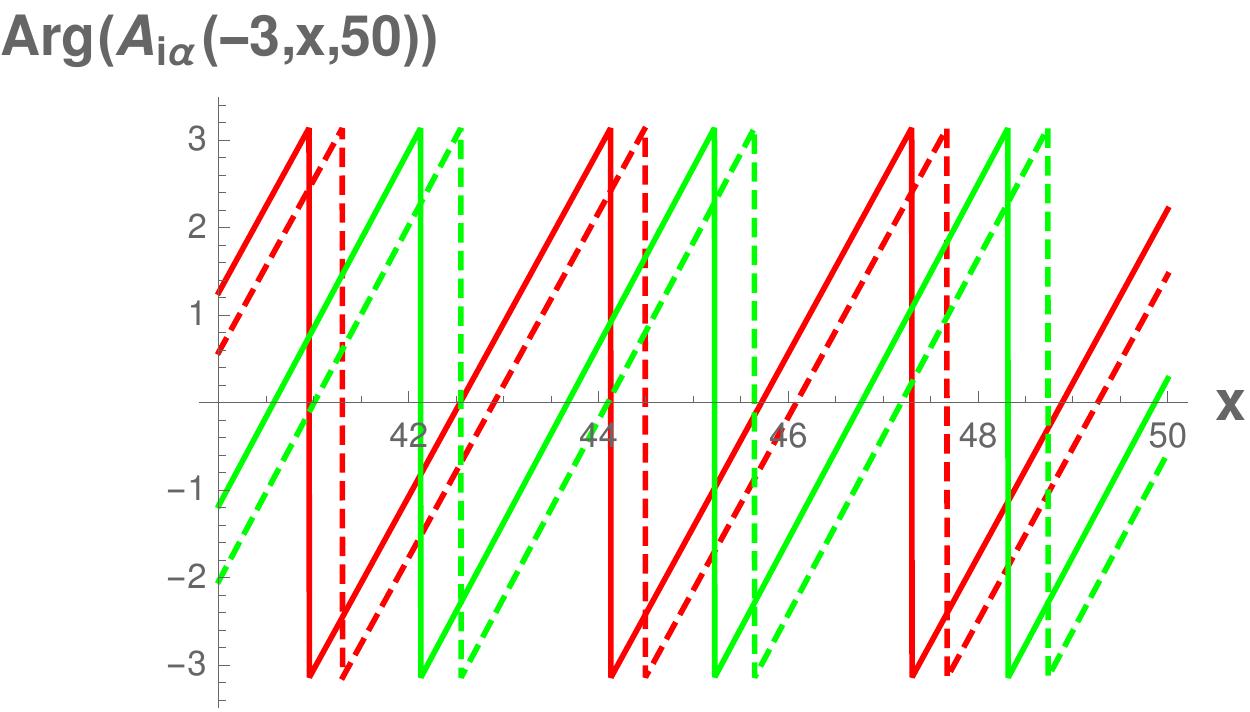}
\caption{The absolute value (left) and phase (right) of $\mathcal{A}_{i\alpha}(p,x,50)$ for $p=-2$ (top) and $p=-3$ (bottom).  The neutrino flavors $i$ are 1 (solid) and 2 (dashed).  The environmental interaction eigenvalues of $0$ (red) and $0.5$ (green) corresponding to $\alpha=0$ and $2$ respectively.  To reduce clutter, the phase is shown over a small range in $x$.}
\label{afig}
\end{center}
\end{figure}

In this subsection we will fix the neutrino mass $m_1$ and the source masses $M_I$ to be
\beq
m_1=0.3 \hsp
M_H=10\hsp
M_L=7.5.
\eeq
We consider only neutrinos whose momenta are equal to
\beq
q_0=2
\eeq
to within an uncertainty $\sigma$.  The initial width squared of the source particle will be
\beq
\beta=1.
\eeq
The energy eigenvalues $\epsilon_\alpha$ of the environmental interactions $H^\prime$ are taken to be
\beq
\epsilon_0=0\hsp
\epsilon_1=0.25\hsp
\epsilon_2=0.5\hsp
\epsilon_3=0.75.
\eeq

Let us begin with a fairly large mass splitting, $m_2=0.4$.  Consider a good momentum resolution $\sigma=0.1$ so that this splitting can have a noticeable effect.  To let each environmental state provide a similar contribution to the probabilities, let us fix
\beq
c_\alpha=2^{3\alpha/2}.
\eeq

At time $t=50$, we plot the amplitudes $\mathcal{A}_{i\alpha}(p,x,50)$ in Fig.~\ref{afig}.  Recoil momenta $p$ of the source particles are set to $p=-2$ and $p=-3$.  As we have assumed that the measured neutrino momentum is equal to $2$, the amplitudes are in general supported at $x>0$.  However the source particle momentum $p+q$ is, within $\sigma$, equal to $0$ and $-1$ when $p=-2$ and $p=-3$ respectively.  Therefore in the later case the $\phi_H$ moved left and so the measured position of the neutrino tends to lower values of $x$.  The phases oscillate quite rapidly, as can be seen, but it is the beating of the phases which leads to neutrino oscillations.  Note that interference is only possible between final states with identical quantum numbers, including the recoil momenta.  Therefore it is the beating at fixed $p$ which yields neutrino oscillations.  On the other hand, one sees that the large environmental energy shifts $\epsilon_\alpha$ considered here have an appreciable effect on the spectra already at $t=50$.  As the environmental state is not measured, the corresponding probabilities $P_\alpha$ will be incoherently summed, degrading the oscillation signal.

\begin{figure} 
\begin{center}
\includegraphics[width=2.5in,height=1.7in]{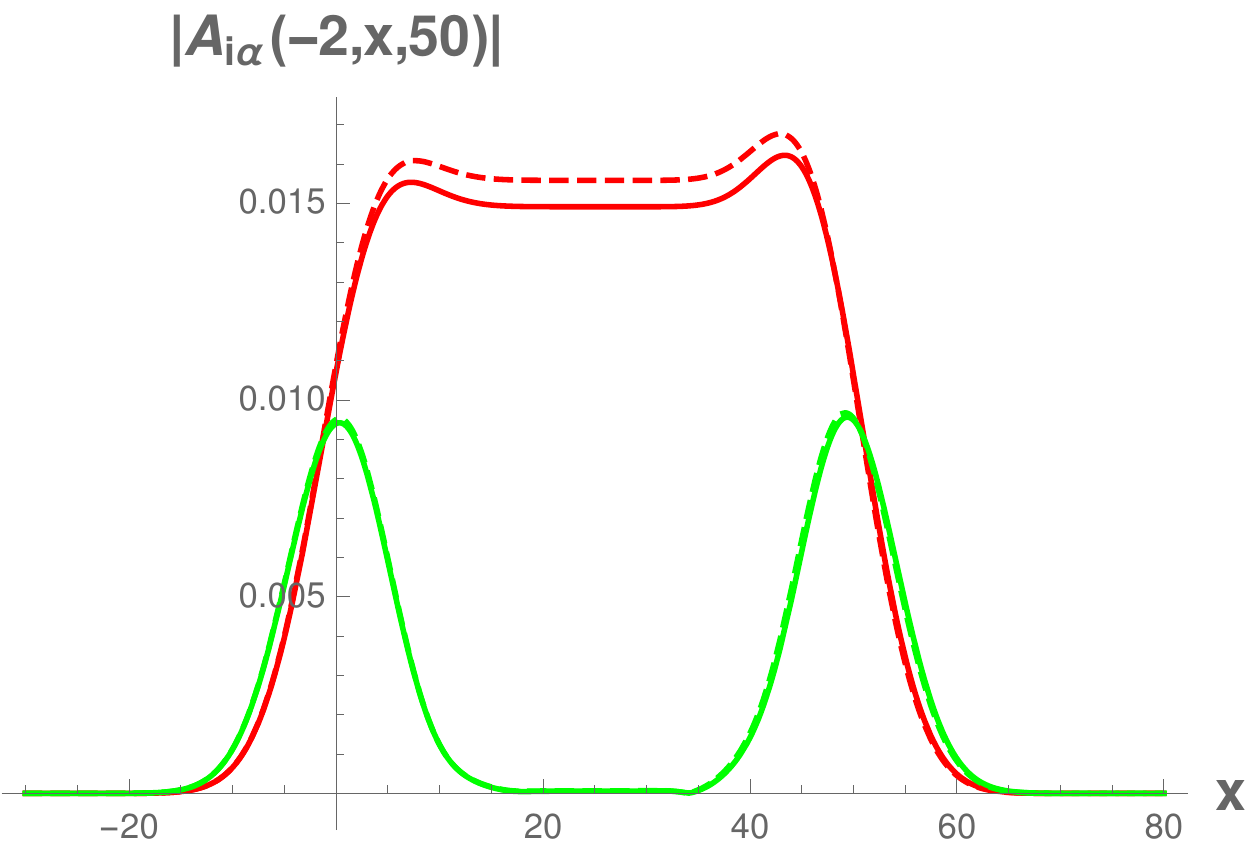}
\includegraphics[width=2.5in,height=1.7in]{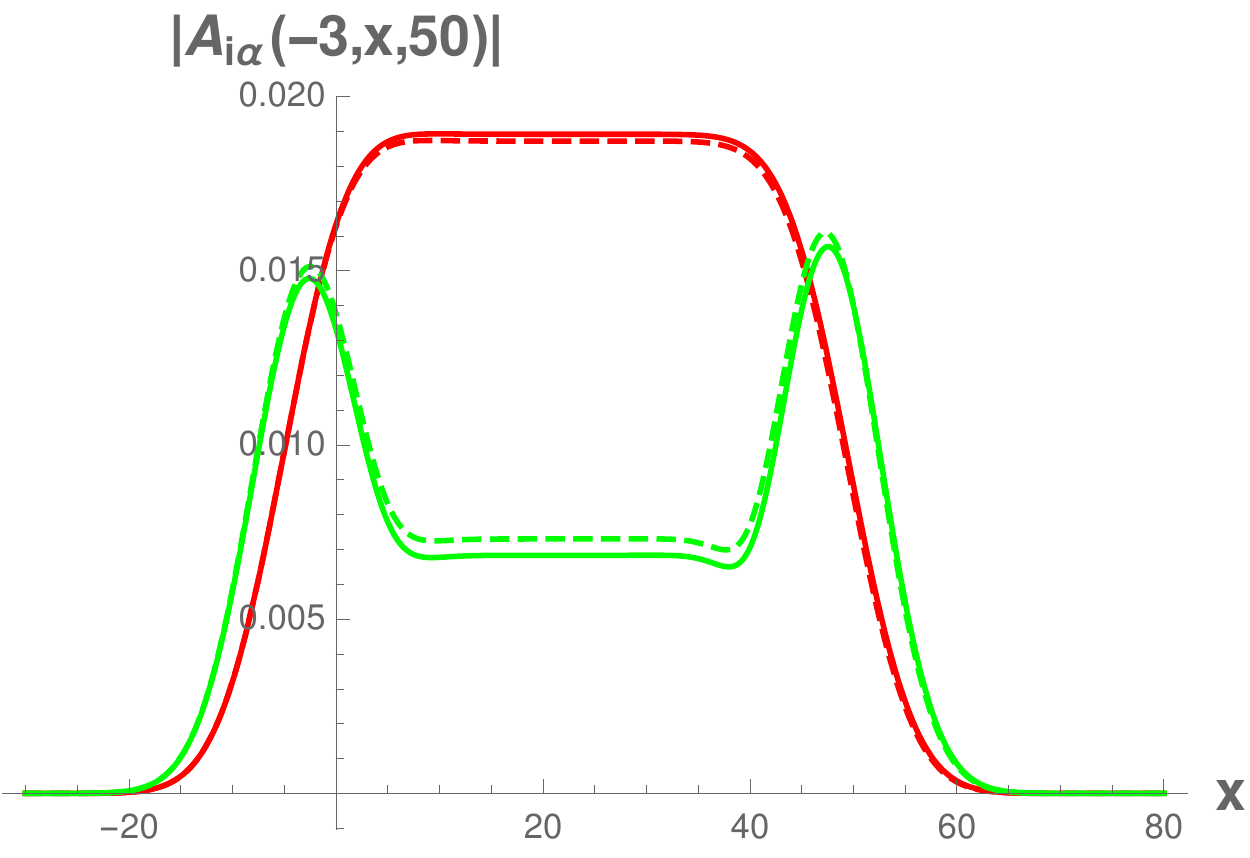}
\caption{As in Fig.~\ref{afig} but for a smaller mass splitting and worse momentum resolution. The phases are not shown.}
\label{a35fig}
\end{center}
\end{figure}

Observe the fairly large fractional difference in the red curves corresponding to the two neutrino flavors in $A_{i0}(-2,x,50)$.  This difference is due to the different phase space for the two masses.  The difference is large because the mass difference is large.  The difference in these amplitudes will damp the neutrino oscillations.   Below, we will see this purely kinetic damping already in the partial probability distributions $P_\alpha$.  Such damping is far too small to be observed at current ultrarelativistic neutrino experiments.

To reduce this purely kinetic source of oscillation damping, we will reduce our mass splitting by setting $m_2=0.35$ and we will worsen our momentum resolution to $\sigma=0.2$ so that the experiment cannot hope to determine the neutrino mass eigenstate from a precise momentum measurement.  To keep the similar contributions to the probabilities, we set
\beq
c_\alpha=2^{3\alpha/4}.
\eeq
At the late times at which oscillations occur.  This has little effect on the phases, so we show the absolute values of the amplitudes for the smaller splitting in Fig.~\ref{a35fig}.  Notice that the difference between the neutrino mass eigenstates is greatly reduced, as expected.  In the case of the environment variable $\alpha=2$, one sees that the amplitude is quite small at intermediate $x$, and in fact vanishingly small at $p=-2$.   This is easy to understand.  Recall that the neutrino momentum is $q=2.0\pm 0.2$.  When $p=2$, then $p+q=0.0\pm 0.2$ and so
\beq
E_{0,2}(p+q)=\sim M_H+\epsilon_2=10.5\hsp
E_{1,i}(p,q)\sim \sqrt{M_L^2+p^2} + q \sim 9.8\pm 0.2
\eeq
 and so the on-shell condition $E_0=E_1$ is only satisfied when the momentum deviates from its measured value at more than the $3\sigma$ level.  Similarly, when $p=3$ one finds
\beq
E_{0,2}(p+q)=\sim \sqrt{M_H^2+1^2}+\epsilon_2=10.55\hsp
E_{1,i}(p,q)\sim \sqrt{M_L^2+p^2} + q \sim 10.1\pm 0.2
\eeq
and so the on-shell condition is excluded at about $2\sigma$.  This explains why the amplitude is small when $p=3$, and very small when $p=2$.  The two peaks in the amplitude at low $x$ and near the light cone are artifacts of the boundary conditions, as in the classical source case considered in Sec.~\ref{classsez}.

\begin{figure} 
\begin{center}
\includegraphics[width=2.5in,height=1.7in]{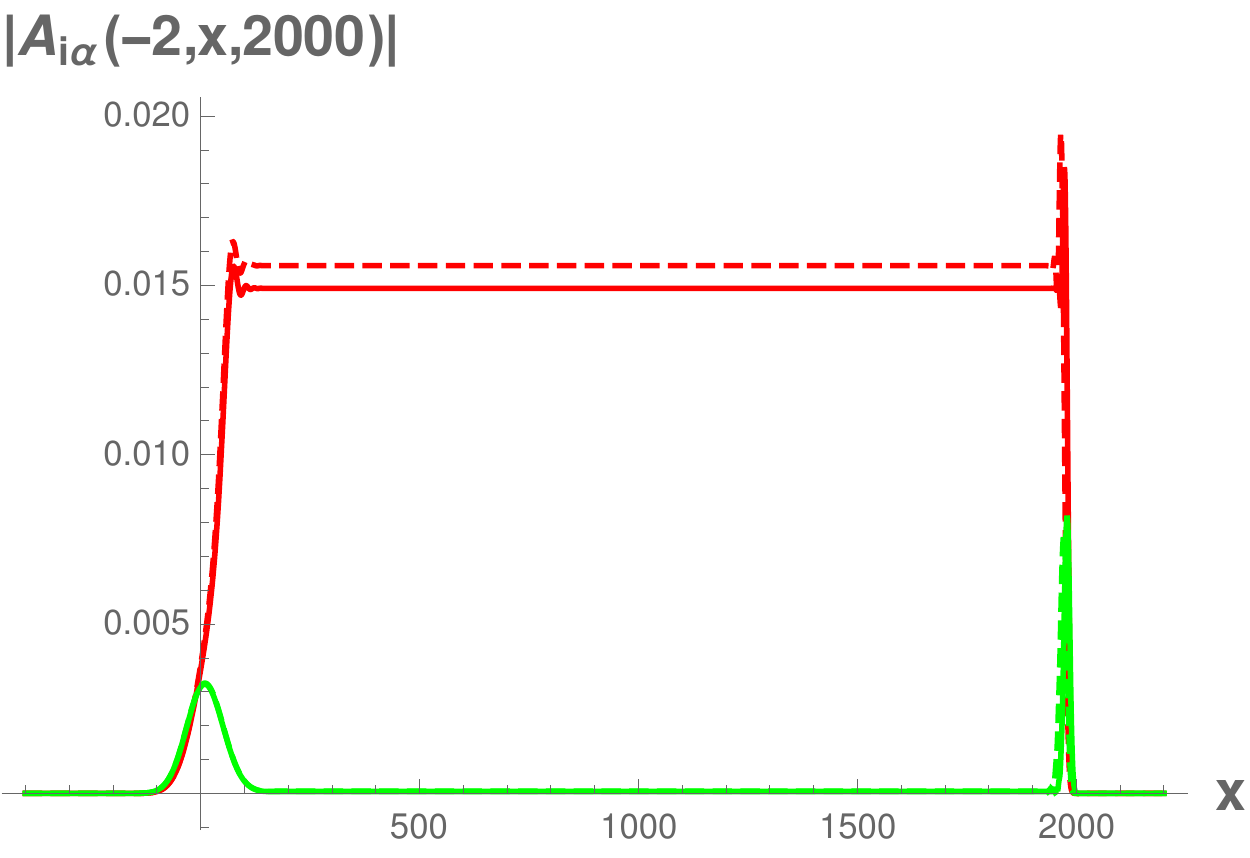}
\includegraphics[width=2.5in,height=1.7in]{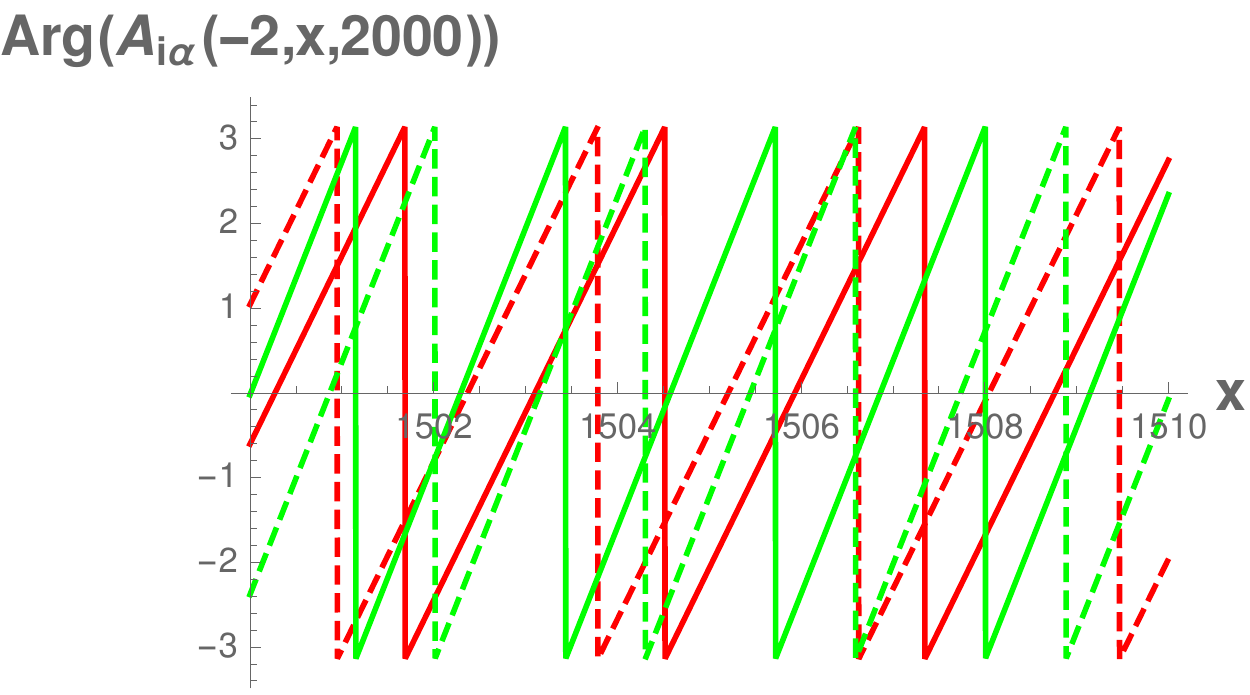}
\includegraphics[width=2.5in,height=1.7in]{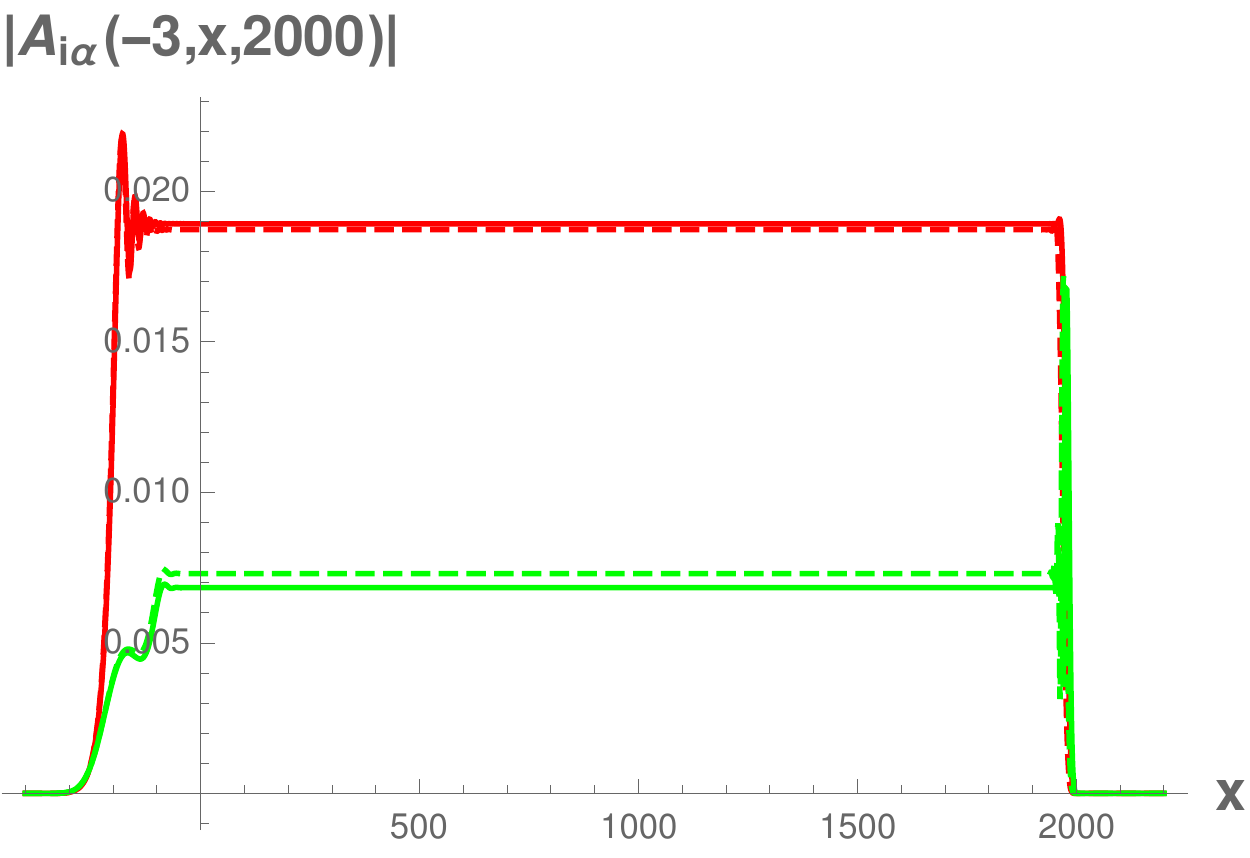}
\includegraphics[width=2.5in,height=1.7in]{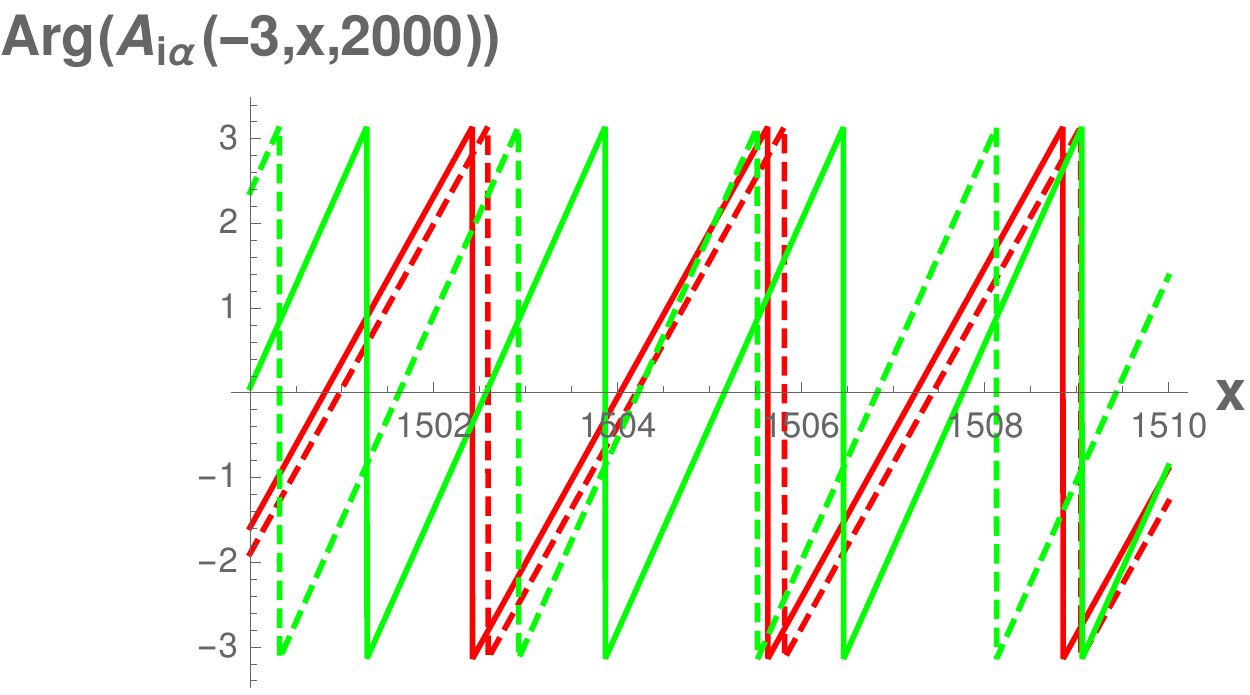}
\caption{As in Fig.~\ref{afig} but at $t=2000$ and for a smaller mass splitting and worse momentum resolution.}
\label{a2000fig}
\end{center}
\end{figure}

At time $t=50$ there are not yet any oscillations and certainly no decoherence.  The amplitudes at $t=2000$ are shown in Fig.~\ref{a2000fig}.  These are qualitatively similar to the $t=50$ case.  However the off-shell contribution at the boundary has become thinner.  Note that while the integral of the off-shell shell region is greatly reduced at later time, as expected, nonetheless in the small region of $x$-space where it is visible due to boundary effects, the amplitudes at $t=50$ and $t=2000$ are similar.

\subsection{Numerical Results: Probabilities}

\begin{figure} 
\begin{center}
\includegraphics[width=2.5in,height=1.7in]{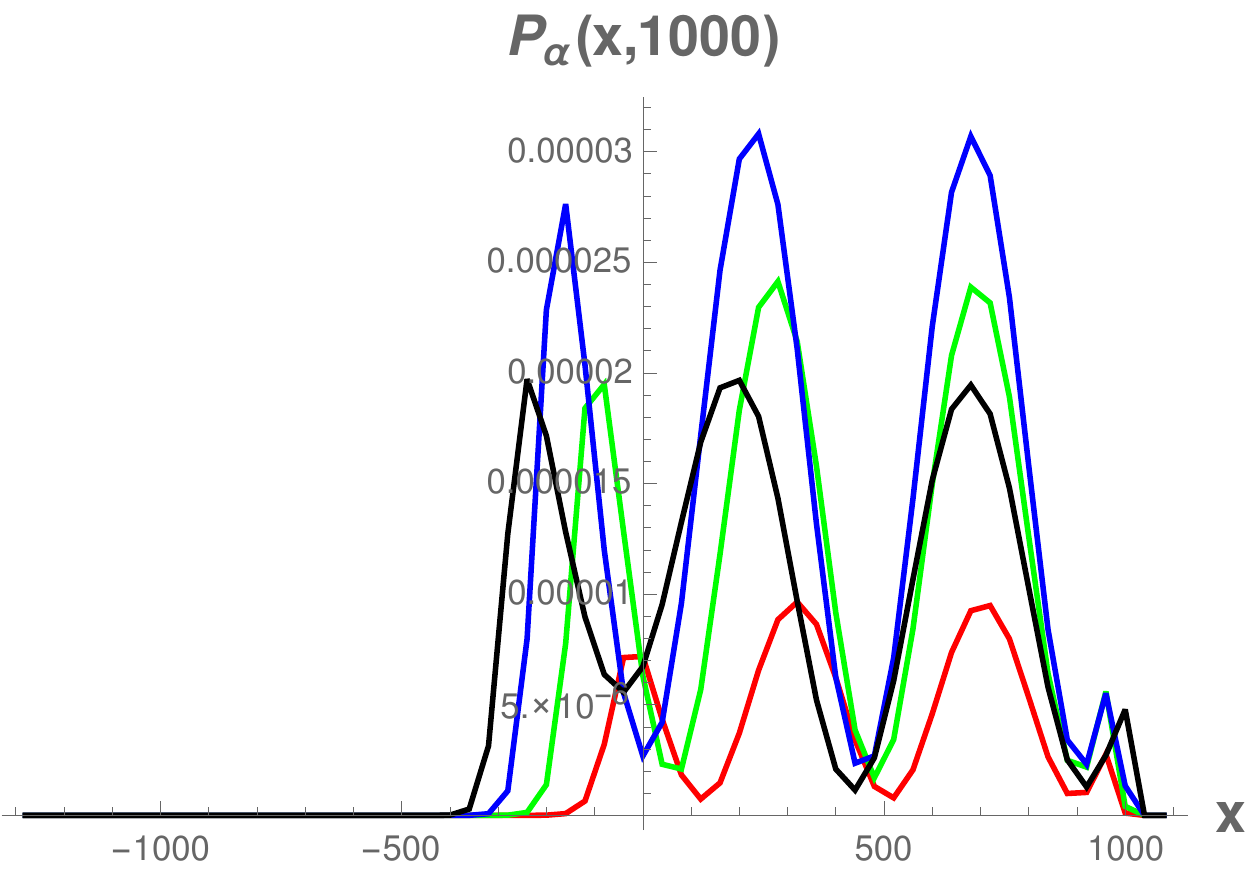}
\includegraphics[width=2.5in,height=1.7in]{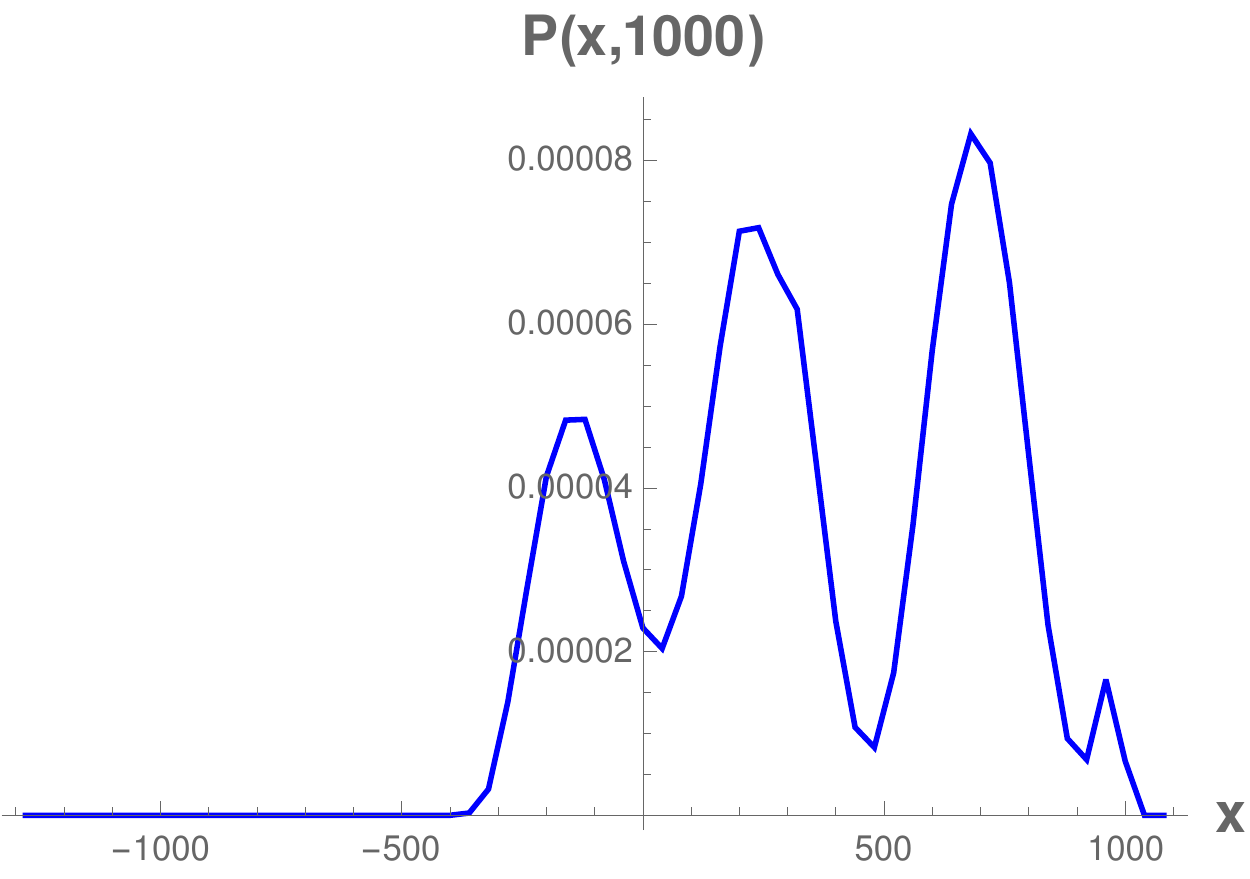}
\includegraphics[width=2.5in,height=1.7in]{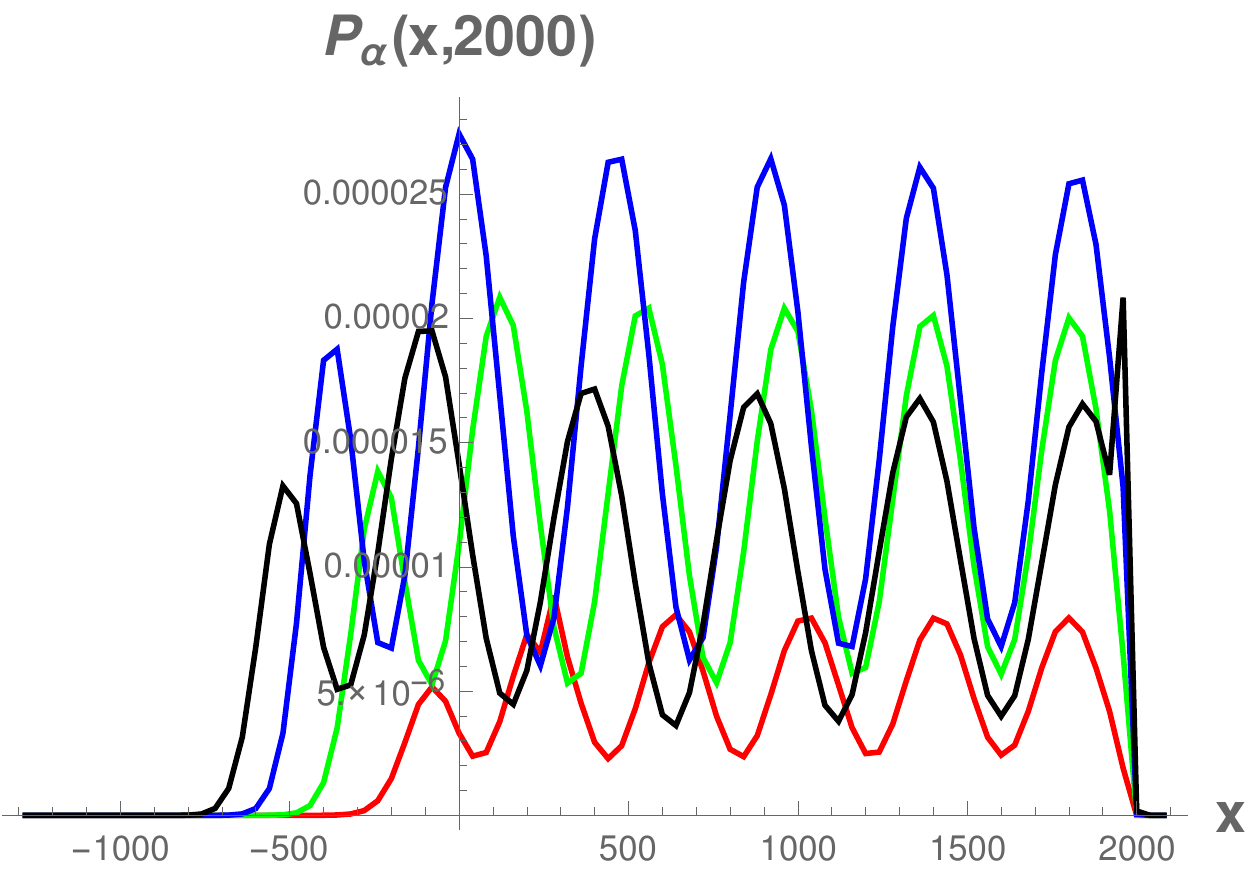}
\includegraphics[width=2.5in,height=1.7in]{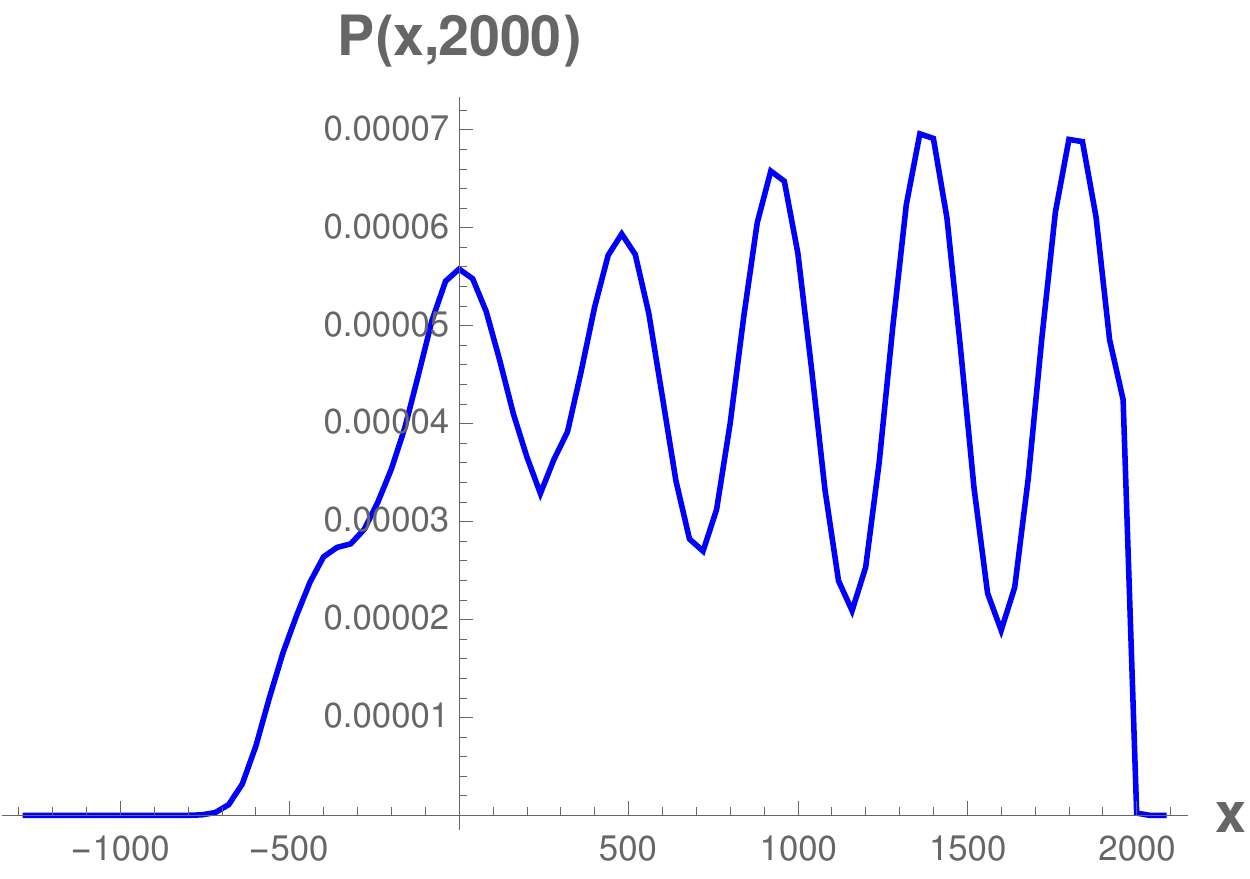}
\includegraphics[width=2.5in,height=1.7in]{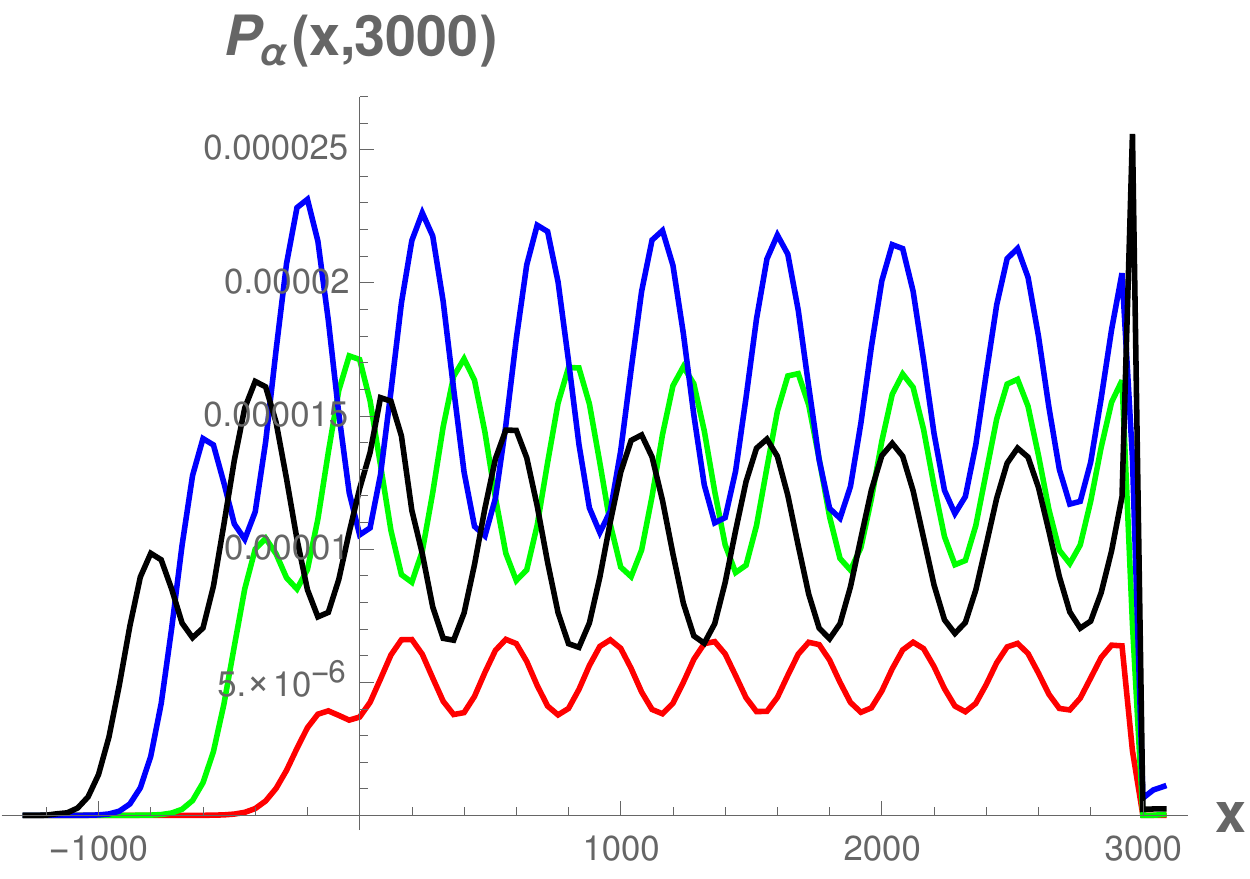}
\includegraphics[width=2.5in,height=1.7in]{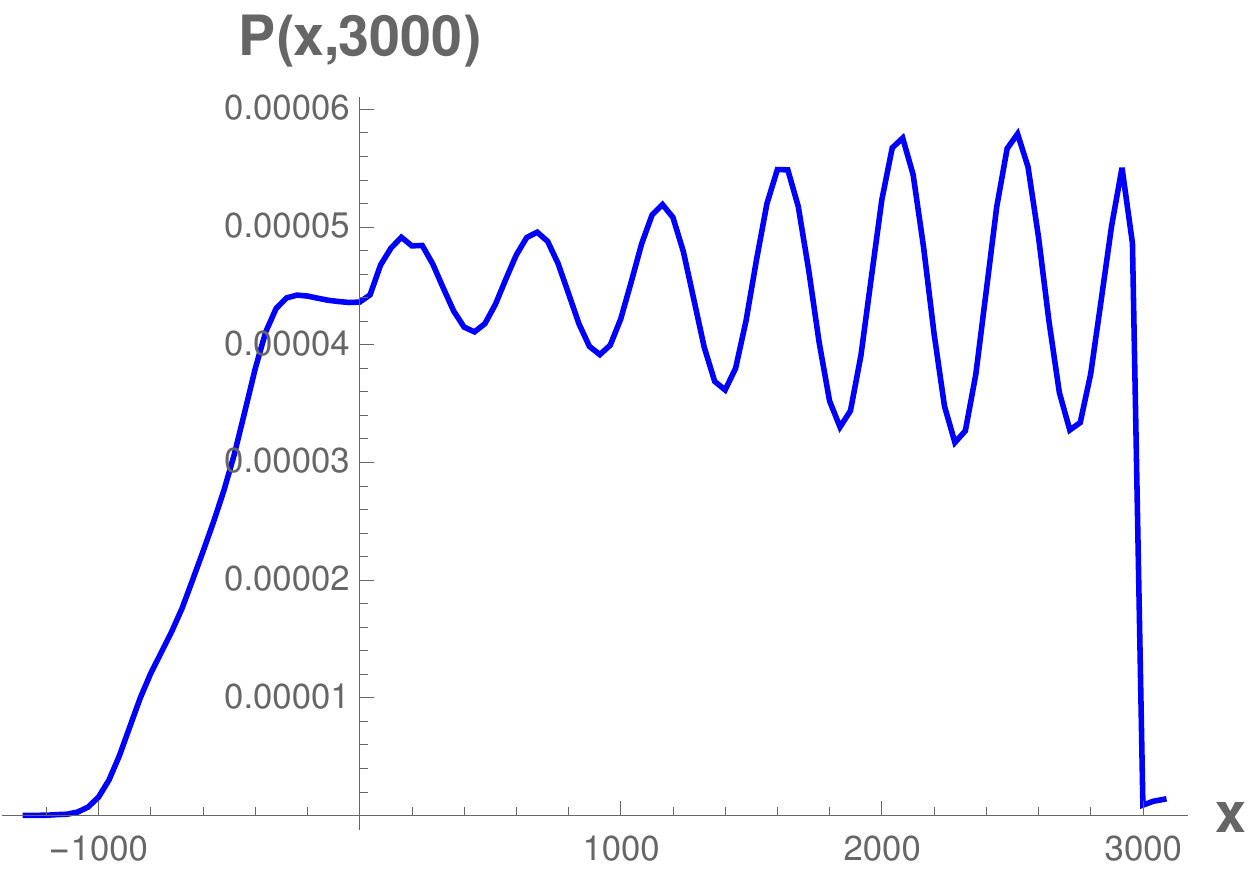}
\caption{The probability densities $P$ (right) and the partial probability densities $P_\alpha$ (left) at $t=1000$ (top), $t=2000$ (middle) and $t=3000$ (bottom).  The environmental interaction energy eigenvalues, for $\Phi_H$, are $\epsilon=0,\ 0.25,\ 0.5$ and $0.75$ corresponding to the red, green, blue and black curves respectively.  Here $m_1=0.3$, $m_2=0.4$ and $\sigma=0.1$.}
\label{pfig}
\end{center}
\end{figure}

Let us return to the large splitting case $m_2=0.4$, $\sigma=0.1$, $c_\alpha=2^{3\alpha/2}$.  The (partial) PDFs are shown in Fig.~\ref{pfig}.  Note that these PDFs are not localized in $x$ as one would expect from wave packets.  This is because all values of $t_0\in[0,t]$ are considered.  If the source particles were measured, this would fix $t_0$ to within some precision and the resulting PDFs would be localized in $x$.  Also a measurement of the neutrino would allow an approximate determination of $t_0$.

The fractional amplitude of the oscillations does appreciably decrease with time, as expected.  However this decrease is mostly present already in this partial probabilities.  It therefore does not result from the environmental interaction, which is not present at all in $P_0(x,t)$.  Rather this is the kinematic decoherence resulting from the fact that the higher mass neutrino has less phase space and so a lower amplitude, as was seen in Fig.~\ref{afig}.

\begin{figure} 
\begin{center}
\includegraphics[width=2.5in,height=1.7in]{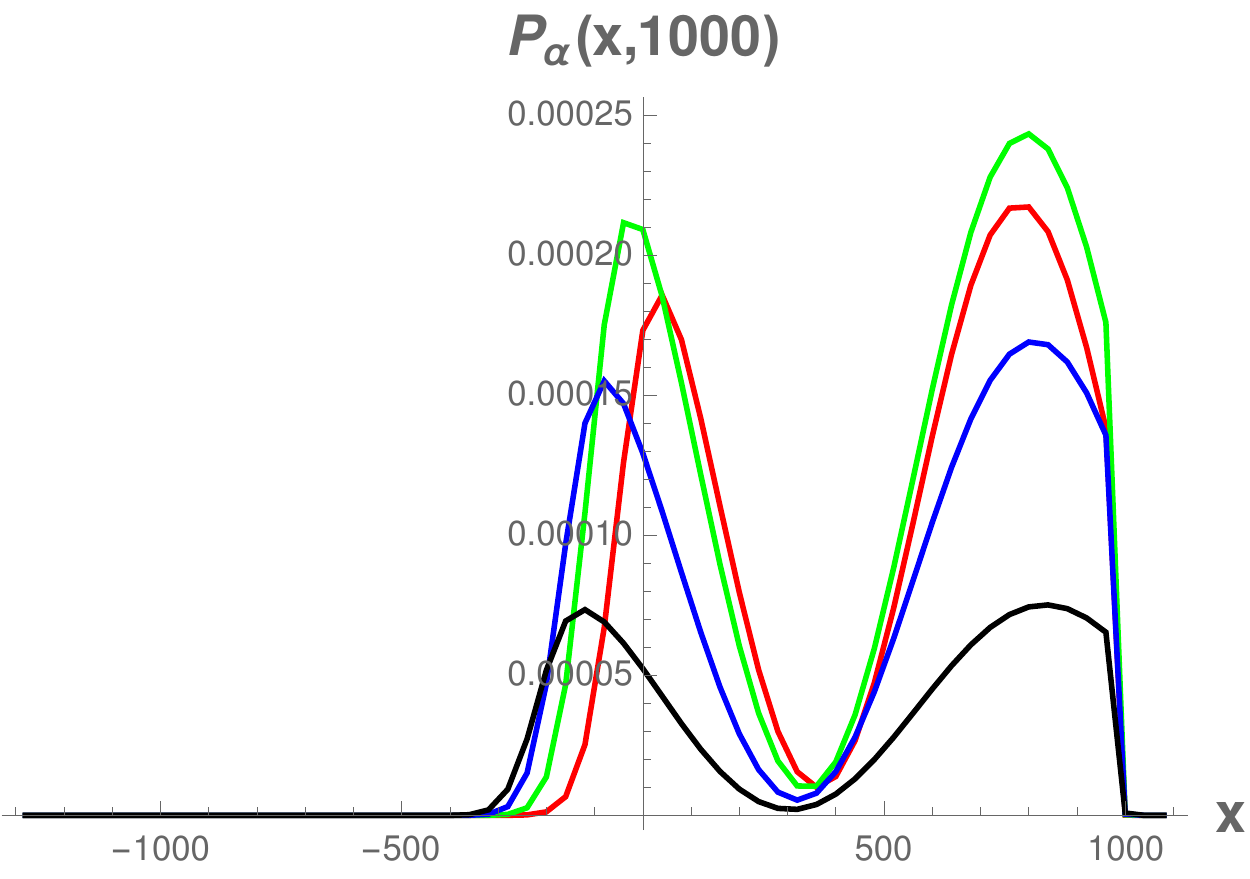}
\includegraphics[width=2.5in,height=1.7in]{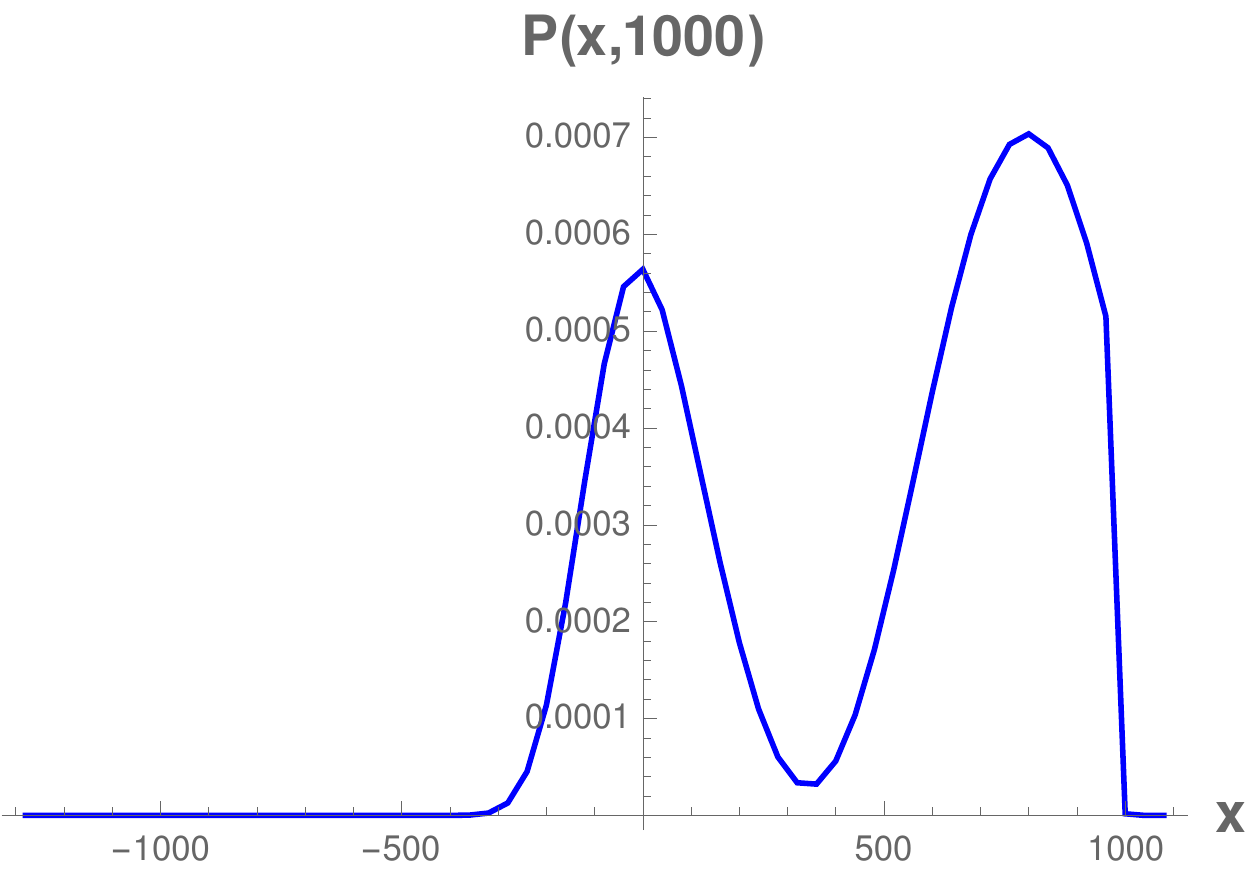}
\includegraphics[width=2.5in,height=1.7in]{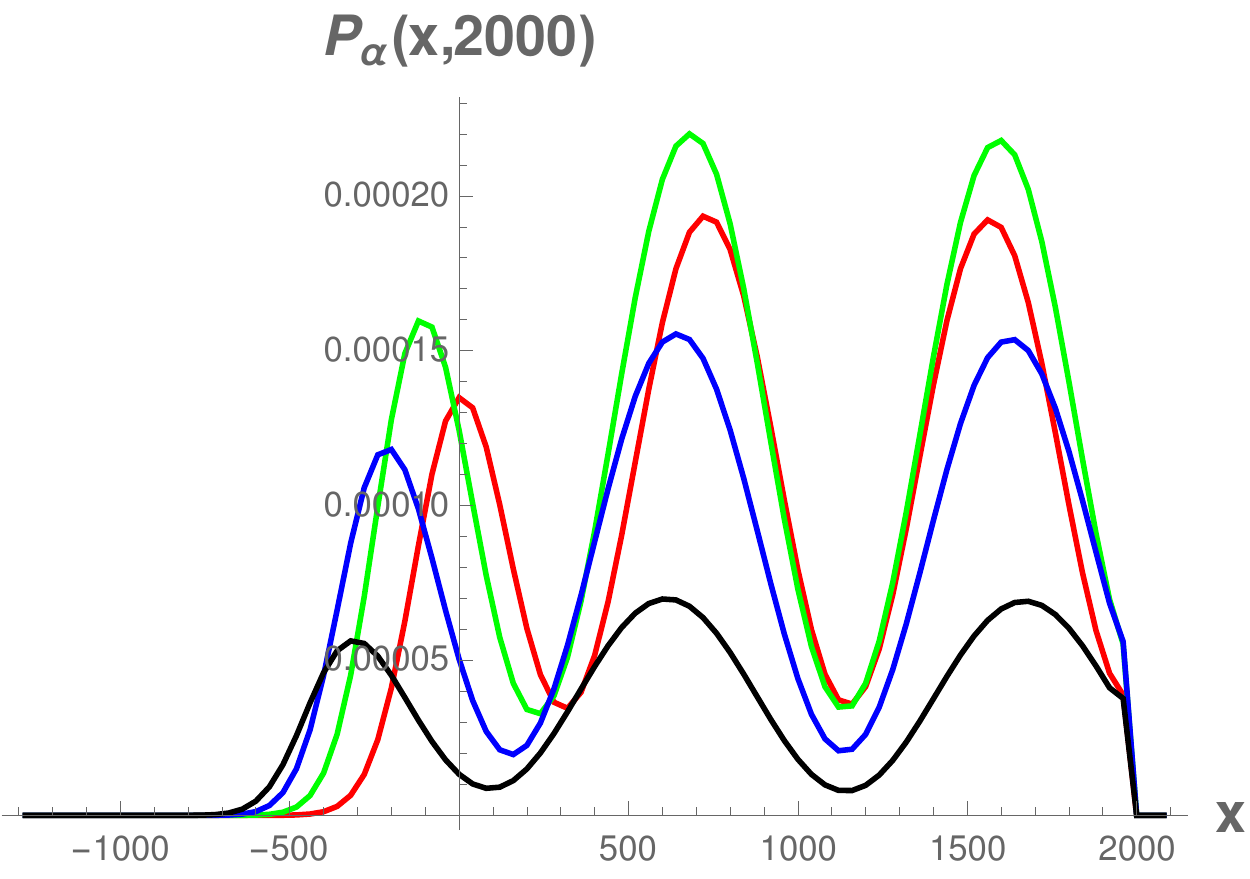}
\includegraphics[width=2.5in,height=1.7in]{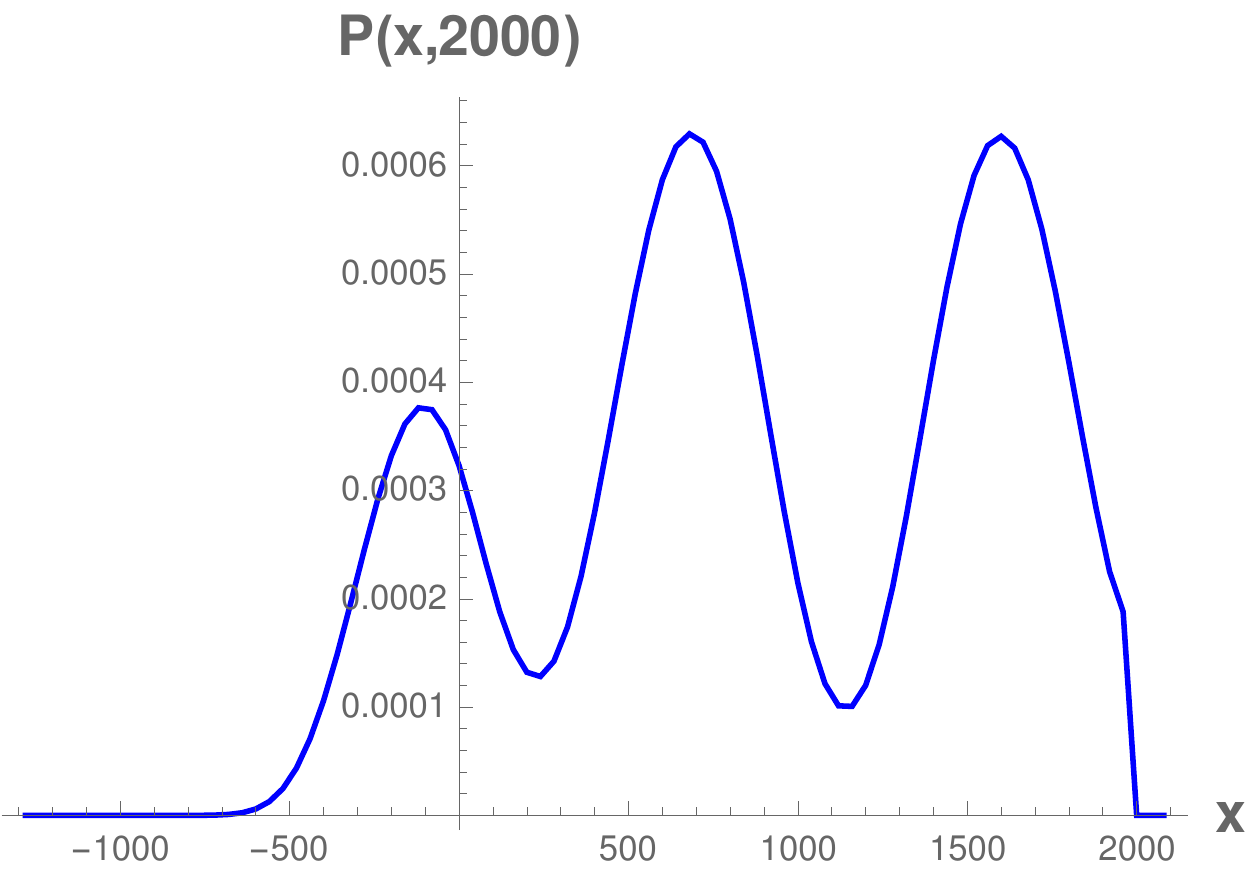}
\includegraphics[width=2.5in,height=1.7in]{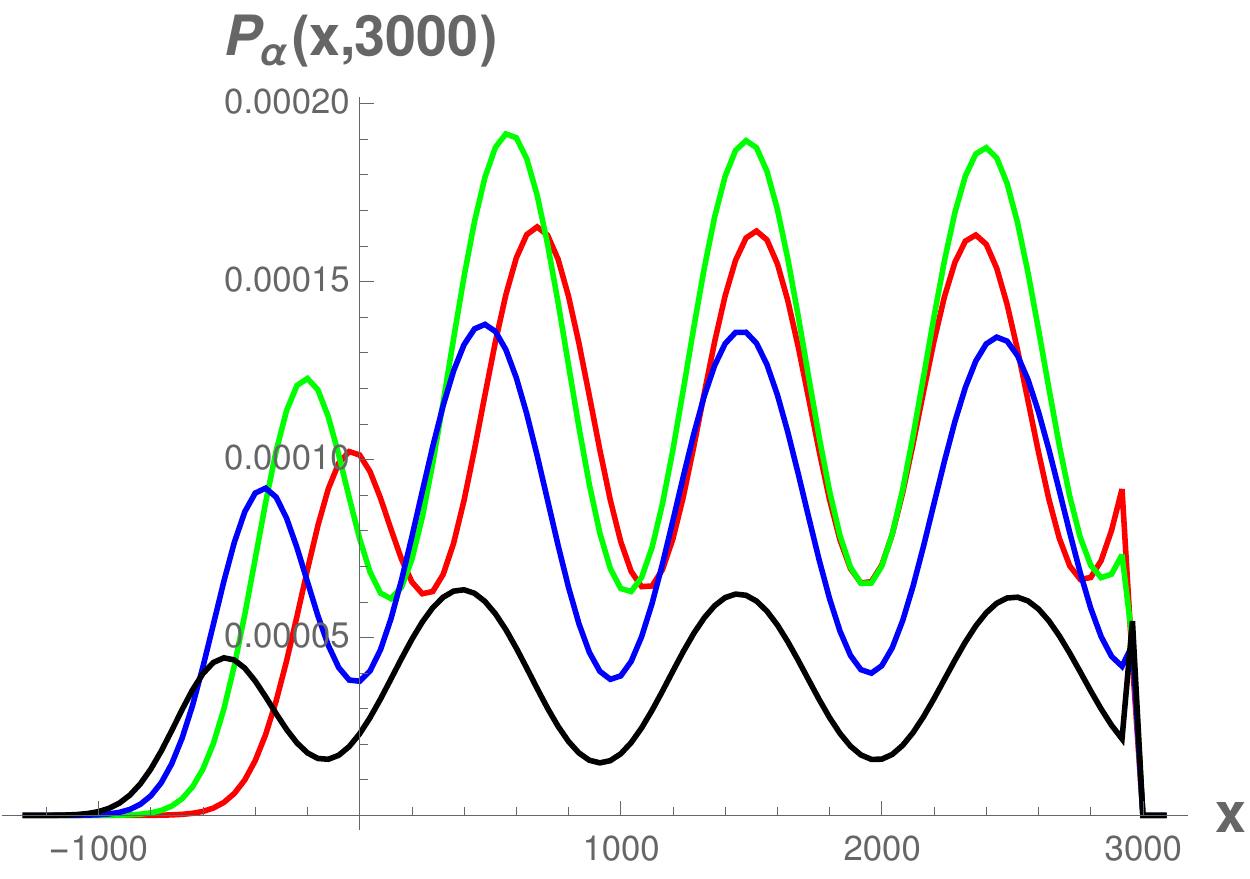}
\includegraphics[width=2.5in,height=1.7in]{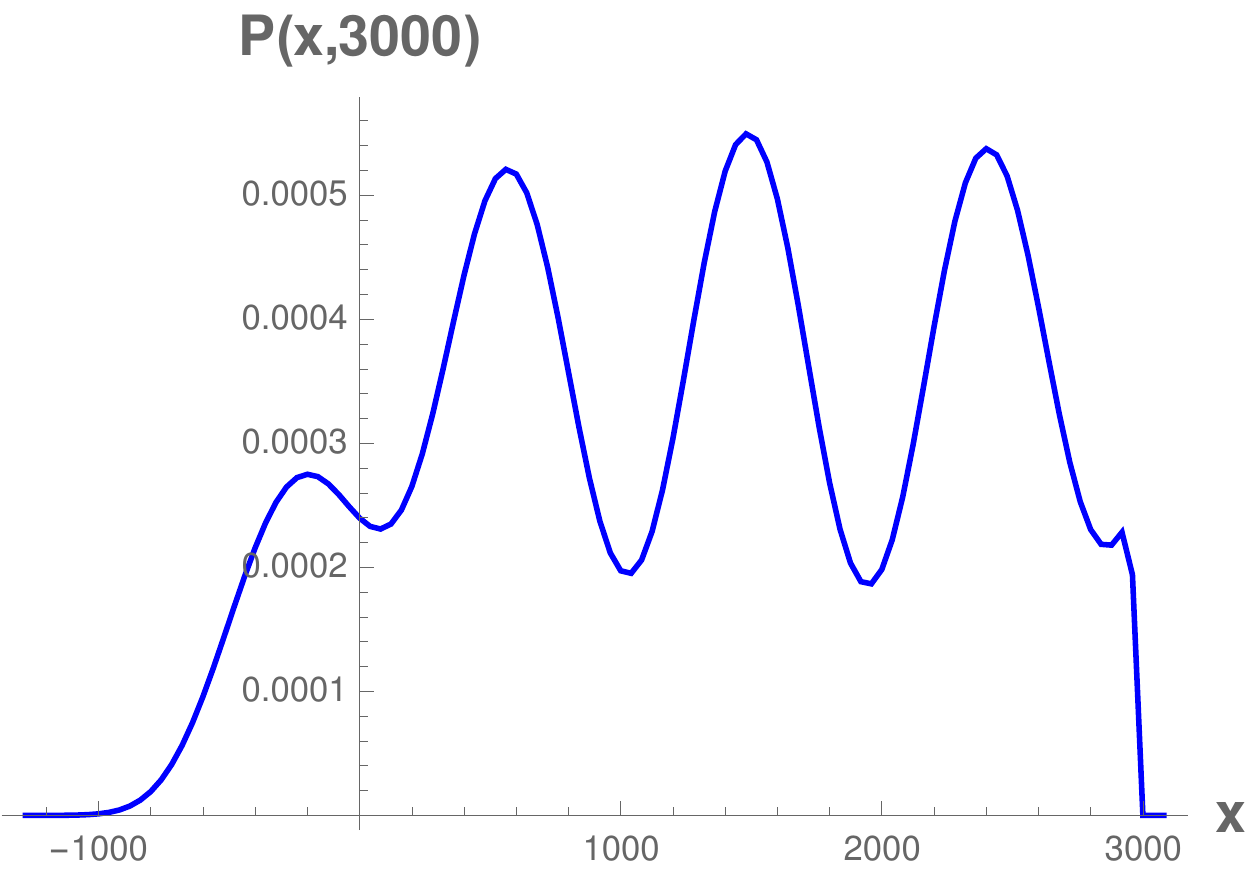}
\caption{The probability densities $P$ (right) and the partial probability densities $P_\alpha$ (left) at $t=1000$ (top), $t=2000$ (middle) and $t=3000$ (bottom).  The environmental interaction energy eigenvalues, for $\Phi_H$, are $\epsilon=0,\ 0.25,\ 0.5$ and $0.75$ corresponding to the red, green, blue and black curves respectively.  Here $m_1=0.3$, $m_2=0.35$ and $\sigma=0.2$.}
\label{p35fig}
\end{center}
\end{figure}

To observe a clear signature of decoherence resulting from environmental interactions, we return to the small splitting case $m_2=0.35$, $\sigma=0.2$, $c_\alpha=2^{3\alpha/4}$.   The corresponding (partial) PDFs are shown in Fig.~\ref{p35fig}.   Now the difference in the amplitudes of the two neutrino mass eigenstates is smaller, as was seen in Fig.~\ref{a2000fig}.   Thus while the amplitude of the partial PDF oscillation does clearly shrink with time, this effect is less pronounced than it was in the large splitting case.   

In both cases one may observe that at lower values of $x$ the oscillation phases differ for the various partial probabilities $P_\alpha$.  By $x\sim 0$ this difference is about $60^\circ$.  Therefore the total probability $P$, which is an incoherent sum of these partial probabilities, has a smaller oscillation amplitude at small $x$ than the partial probabilities.  This is the decoherence arising from destructive interference between the various environmental interaction eigenstates.  One may observe in Fig.~\ref{p35fig} that by $x\sim 0$, at $t=3000$, it nearly removes the oscillation minimum.

\begin{figure} 
\begin{center}
\includegraphics[width=2.5in,height=1.7in]{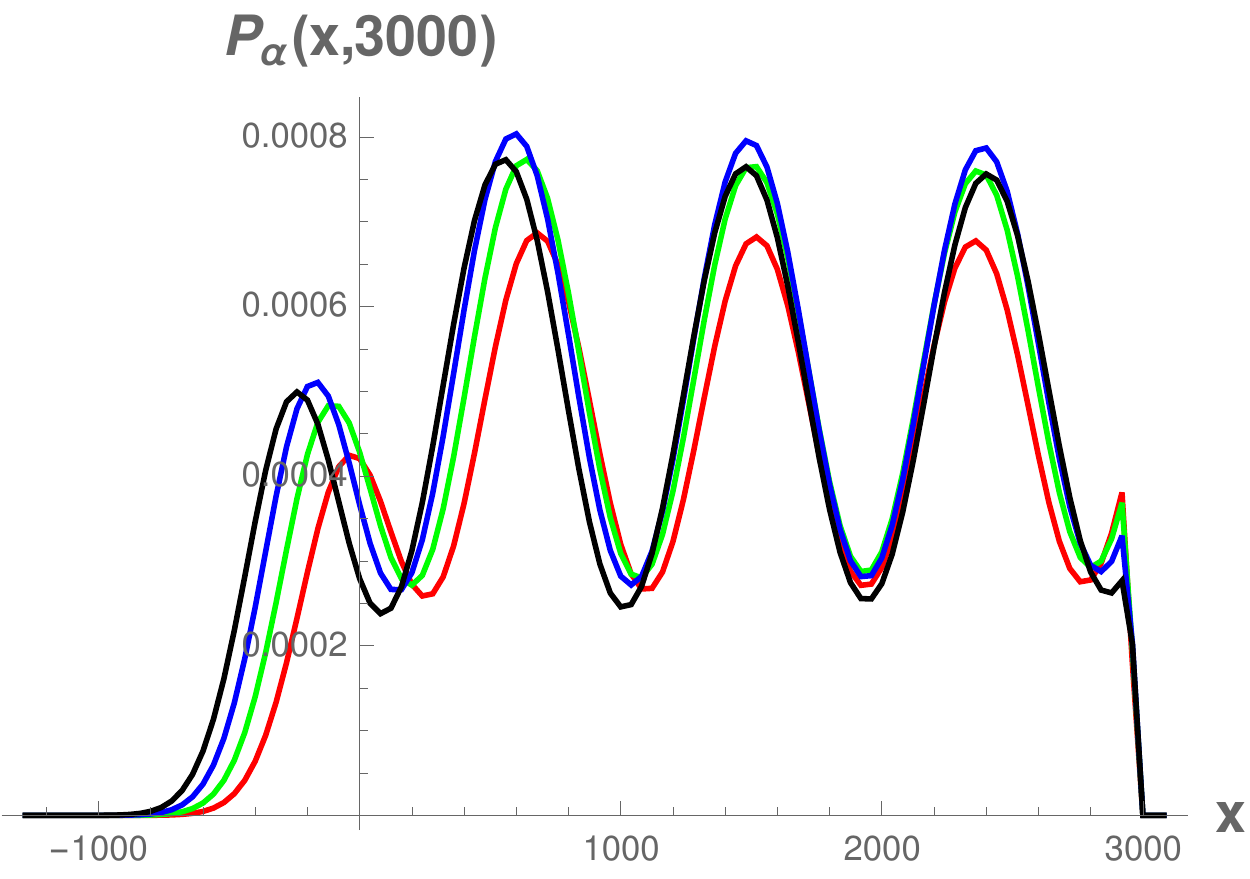}
\includegraphics[width=2.5in,height=1.7in]{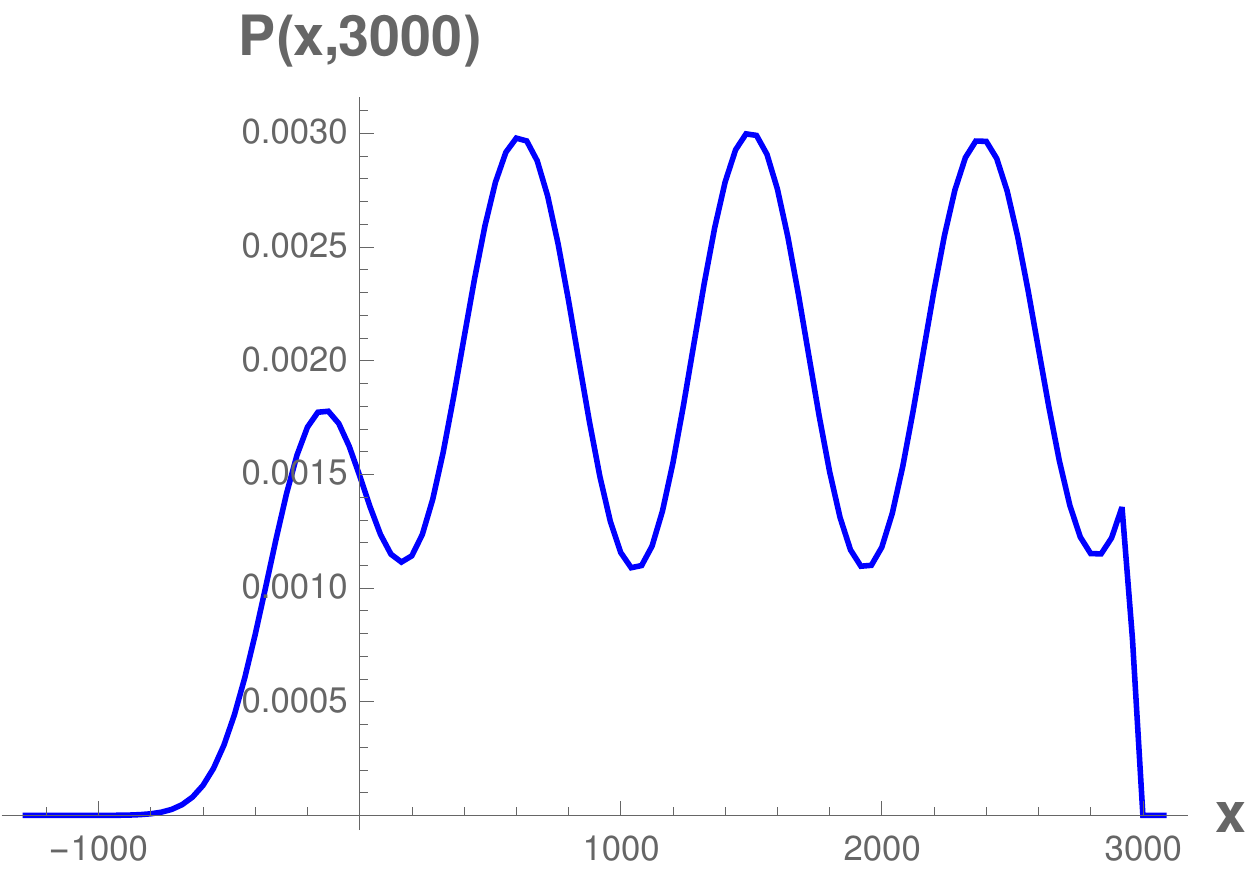}
\caption{The probability densities $P$ (right) and the partial probability densities $P_\alpha$ (left) at  $t=3000$.  The environmental interaction energy eigenvalues, for $\Phi_H$, are $\epsilon=0,\ 0.1,\ 0.2$ and $0.3$ corresponding to the red, green, blue and black curves respectively.  Here $m_1=0.3$, $m_2=0.35$ and $\sigma=0.2$.}
\label{p10fig}
\end{center}
\end{figure}

As one might expect, if the environmental interaction is weakened then so is the interference.  In Fig.~\ref{p10fig} we reduce the environmental interaction to 
\beq
\epsilon_0=0\hsp
\epsilon_1=0.1\hsp
\epsilon_2=0.2\hsp
\epsilon_3=0.3\hsp
c_\alpha=2^{3\alpha/10}.
\eeq
One can see that the various partial probabilities $P_\alpha$ oscillate with little phase difference and so constructively interfere.  In this note we will not systematically study the necessary environmental interaction $\epsilon$ for decoherence to set in at a fixed time $t$.  However in this example our results appear to be consistent with the thesis that for the first few oscillations $\epsilon$ should be of the same order as the neutrino momentum.  It is also clear that decoherence has a large effect on the positions where the neutrinos have oscillated more times.  In our figures this corresponds to the low values of $x$, but at JUNO it would correspond to the lower energy part of the spectrum.

\section{Conclusions}

In this note we have introduced a simple model of neutrino production, oscillation and decoherence due to environmental interactions of the source particle.  This model was treated consistently in quantum field theory and is sufficiently simple that the various wave functions have been calculated explicitly, albeit numerically.  Interactions between the source particle(s) and the environment yield a characteristic coherence time.  The usual approach is to consider a Gaussian neutrino wave packet with width equal to this coherence time but then to neglect the entanglement with the environment, and often also the entanglement with the source.  Following in the suggestion of \cite{cgl}, our approach is different.  We have kept the full entangled state consisting of the neutrino, source particle and also the environment.    Our first principles calculation of the neutrino wave function can be used to test various conjectures in literature, such as the covariant wave packet conjecture of Refs.~\cite{naumov1,naumov2}.  We have not yet included a model of measurement, but to do so in the future will be straightforward.  A consistent treatment of entanglement and measurement will allow us to test the revival mechanism of Refs.~\cite{revival,mcdonald}. 

We have worked in a basis in which the environmental interactions $H^\prime$ are diagonal.  As the Hamiltonian is Hermitian, it may always be diagonalized in principle.  While in the case of accelerator neutrinos, the interactions may be relatively simple \cite{accdec} and so such a diagonalization is straightforward, in the case of reactor neutrinos there are a number of distinct interactions contributing to $H^\prime$ and an explicit diagonalization would be difficult.  However, our analysis suggests that the environmental interaction is appreciable only if the eigenvalues of $\epsilon_\alpha$ are not too far beneath the neutrino energy, or perhaps the neutrino energy divided by the number of oscillations.  In the case of reactor neutrinos, interactions within the nucleus itself after a $\beta$ decay may be expected to have characteristic energies of hundreds of keV, which would be sufficient.  The inner electrons have binding energies of 10s of keV, and so interactions with these electrons may also cause noticeable coherence, at least in experiments such as JUNO that are sensitive to many oscillations.  On the other hand interatomic interactions, which are commonly used to set the coherence scale \cite{rich,boriserr}, have energy scales of eV, and so are unlikely to have noticeable decoherence effects in any proposed reactor neutrino experiment.   We have seen that only the difference between the interaction strength before and after the neutrino emission contributes to decoherence, further reducing the impact of interatomic interactions.

\section* {Acknowledgement}

\noindent
We are greatful to Carlo Giunti for comments on this draft.  JE is supported by the CAS Key Research Program of Frontier Sciences grant QYZDY-SSW-SLH006 and the NSFC MianShang grants 11875296 and 11675223.  EC is supported by NSFC Grant No. 11605247, and by the Chinese Academy of Sciences Presidents International Fellowship Initiative Grant No. 2015PM063.  JE and EC also thank the Recruitment Program of High-end Foreign Experts for support.

\end{document}